\documentclass[11pt,a4paper,fleqn]{article}
\usepackage[utf8] {inputenc}
\usepackage{authblk}
\usepackage{geometry}
\usepackage{xcolor}
\usepackage{natbib}
\usepackage{graphicx}
\usepackage{pdflscape}
\usepackage{appendix}
\usepackage{setspace}
\usepackage{makecell}
\usepackage{graphbox}
\usepackage{soul}

\usepackage{colortbl}

\usepackage{threeparttable}
\usepackage{booktabs}

\usepackage{amssymb}
\usepackage{amsbsy}
\usepackage{bm}
\usepackage{amsmath}
\usepackage{amsthm}
\usepackage{graphicx}
\usepackage{mathtools}

\usepackage{enumitem}
\newlist{steps}{enumerate}{1}
\setlist[steps, 1]{label = Step \arabic*:}

\usepackage{dcolumn}
\newcolumntype{d}[1]{D{.}{.}{#1}}



\usepackage{caption}
\usepackage{subcaption}
\usepackage{forest}
\definecolor{nblue}{HTML}{000660}
\usepackage[colorlinks=true,urlcolor=nblue,linkcolor=nblue,citecolor=nblue]{hyperref}
\title{\LARGE \textbf{Asymmetries in International Financial Spillovers}\thanks{
\noindent Corresponding author:  Karin Klieber, Monetary Policy Division, Oesterreichische Nationalbank. \textit{Address}: Otto-Wagner-Platz 3, 1090 Vienna, Austria. \textit{Email}: \href{mailto:karin.klieber@oenb.at}{karin.klieber@oenb.at}. The views expressed in this paper do not necessarily reflect those of the European Commission, the European Central Bank, the Oesterreichische Nationalbank or the Eurosystem.}

\vspace{0.3em}}
\author[1]{Florian \MakeUppercase{Huber}}
\author[2]{Karin \MakeUppercase{Klieber}}
\author[3]{Massimiliano \MakeUppercase{Marcellino}}
\author[4,6]{Luca \MakeUppercase{Onorante}}
\author[5]{Michael \MakeUppercase{Pfarrhofer}}
\affil[1]{\textit{University of Salzburg}}
\affil[2]{\textit{Oesterreichische Nationalbank}}
\affil[3]{\textit{Bocconi University, Baffi, BIDSA, CEPR, IGIER}}
\affil[4]{\textit{European Commission (Joint Research Center)}}
\affil[5]{\textit{Vienna University of Economics and Business}}
\affil[6]{\textit{European Central Bank}}

\begin{document}

\maketitle\thispagestyle{empty}\normalsize\vspace*{-2em}\small

\begin{center}
\begin{minipage}{0.8\textwidth}

\noindent 
We examine asymmetries in the international transmission of US financial shocks. We develop a nonlinear factor VAR in which structural shocks enter through monotone Bayesian additive regression tree transmission functions. Sign, zero, and anchor restrictions identify five shocks; convexity constrains the financial shock’s impact function. Using data for the United States, euro area, United Kingdom, Japan, and Canada, we find that adverse financial shocks have larger effects than equally sized benign shocks, with the gap widening as shock size increases. Asymmetry spans output, prices, interest rates, and equity prices. A linear benchmark understates adverse effects and overstates benign effects abroad.


\end{minipage}
\end{center}

\setstretch{1.5}\normalsize\renewcommand{\thepage}{\arabic{page}}

\newpage
\section{Introduction}
Financial shocks have large, asymmetric and international effects on macroeconomic aggregates \citep[see, e.g.,][]{dovern2014international, ciccarelli2016commonalities, prieto2016time, gerba2024inspecting}. Domestically, disruptions to credit supply tighten borrowing constraints, force deleveraging and depress output and investment. Credit spreads also carry information about the business cycle beyond that contained in standard macroeconomic aggregates \citep{gilchrist2012ebp, he2013intermediary, christiano2014risk}. Internationally, risk premia, credit spreads, capital flows and asset prices co-move across countries. These co-movements characterize the global financial cycle, with financial conditions tending to tighten in periods of stress \citep{rey2013dilemma, miranda2020us, habib2019global}. Policymakers in central banks and governmental institutions aim to smooth business cycles and to alleviate the effects of such disruptions. This requires an understanding of how financial shocks affect the economy and propagate across borders.

Two features of international propagation are well documented. The first is that much of it is driven by conditions at the center of the international monetary system. US-based financial shocks depress activity abroad, and the effects may vary over time \citep[see][]{eickmeier2015, abbate2016changing, cesa2022financial}. The global financial crisis is a prominent example. The second is that transmission is nonlinear in the sign and the size of the shock. Contractionary financial shocks have a much stronger effect than benign ones of the same magnitude \citep[see][]{barnichon2022effects, mumtaz2022impulse, forni2024nonlinear}. The same holds for monetary policy, where the effects depend on the sign of the shock, on its size, and on the state of the economy \citep{tenreyro2016pushing, barnichon2018functional, ascari2022sign}. Both features matter for policy. An institution outside the US that assesses its exposure to US financial conditions needs to know more than the average spillover. It needs to know whether an adverse disturbance transmits more strongly than a benign one, and whether a large disturbance transmits more than proportionally.

Most existing models examine either international spillovers or nonlinear transmission. The multi-country literature that measures spillovers is predominantly linear. Panel VARs typically feature time-invariant coefficients \citep{canova2013panel, koop2016model}, and global VAR and factor-augmented VAR treatments of US shocks are linear as well \citep{feldkircher2016international, georgiadis2017bi, bai2022macroeconomic}. These models force the response to a benign shock to be the mirror image of the response to an adverse shock of equal size, and the response scales proportionally with the size of the shock. The nonlinear literature that can accommodate asymmetry rests mostly on closed-economy models \citep{barnichon2022effects, mumtaz2022impulse, forni2024nonlinear}. It therefore provides limited evidence on nonlinear international transmission. Time-varying parameter, threshold and Markov switching specifications relax linearity in the propagation mechanism \citep{canova2009estimating, hubrich2013financial, billio2016interconnections}. However, in specifications with a linear impact equation, the impact response remains proportional to shock size conditional on the date or regime. Work on asymmetries in the spillovers of financial shocks is accordingly sparse, with \citet{abbate2016changing} providing an exception. Two modeling difficulties help explain this gap. The first is dimensionality. Even in the two-country case the number of time series doubles, and adding more countries or more domestic series increases the number of parameters further. The second is that nonlinear modeling requires the researcher to commit to a specific form of the nonlinearity, which can be difficult to justify in a multi-country system.

Our first contribution is evidence on asymmetries in the spillovers of US-based financial shocks to four major economies, namely the euro area (EA), the United Kingdom (UK), Japan (JP) and Canada (CA). We develop a nonlinear factor-augmented multi-country VAR based on the approach of \cite{clark2026flexible}. We extend this model by allowing each structural shock to enter the observation equation through a nonlinear function of itself, which we call its transmission function. We model each transmission function with monotone Bayesian additive regression trees (BART) \citep{chipman2010bart,  chipman2022mbart}.


The model addresses both difficulties. First, the factor structure allows for more than one country while keeping the dimension of the model under control. The observation equation combines a linear VAR component, a flexible nonlinear common component built from BART, and a small block of structurally interpretable factors. The BART component absorbs nonlinear common variation that is not structurally interpretable, such as state dependence in the propagation mechanism or threshold effects, without requiring these features to be specified in advance. Second, we allow the impact of each structural shock to depend nonlinearly on its size through a univariate transmission function. The shape of this function determines how the impact response varies with the sign and size of the shock. Since the function is low-dimensional, we restrict its shape a priori. Monotone trees leave the number and location of its kinks to the data \citep{buckmann2025blockwise}.

A third difficulty is the COVID-19 period. In the spring of 2020 many of our series moved by far more than at any other point in the sample. Left untreated, a few such observations can dominate the estimates of a flexible model. Its nonlinear components can then fit the collapse of output in April 2020, and large shocks replay that collapse in the impulse responses. We therefore let the measurement errors share a common volatility scale that is free in March, April and May 2020 and decays afterwards, in the spirit of \citet{lenza2022vector}.

Our second contribution is the identification scheme. We do not identify the financial shock through a recursive ordering since financial variables and macroeconomic aggregates interact within the period. This makes separating credit supply disturbances from demand shocks or from policy difficult, requiring more than timing assumptions based on zero restrictions \citep{caldara2016macroeconomic, furlanetto2019identification, cesa2022financial}. Nonlinear impact effects introduce an additional identification problem. The tools built for linear structural VARs do not carry over unchanged. We thus identify the shock through three layers of restrictions. The first layer is a set of sign restrictions on the impact loadings. They identify five structural disturbances, namely supply, demand, monetary policy, investment and financial shocks, following \cite{furlanetto2019identification}. A point restriction per shock, which we call an anchor, fixes one loading and normalizes the scale of its transmission function. We also assume that industrial production in the four foreign economies cannot respond to a US financial shock within the month. The second layer requires every transmission function subject to a sign restriction to be non-decreasing. Without it, a flexible function can slope downward for some values of the shock. The impact response then takes the opposite sign to the loading, so the sign restriction is violated even though the loading satisfies it. A non-decreasing transmission function ensures that the impact response respects the sign restriction on the loading, allowing for a zero response.\footnote{This is a general point about sign identification in nonlinear models, and it is not specific to our framework.} The third layer requires the financial transmission function to be weakly convex as well as non-decreasing. The asymmetry documented by the credit shock literature thus becomes an identifying restriction on the impact response. The linear case remains admissible. Convexity restricts a shape and not a magnitude, so the degree of asymmetry is estimated and not imposed. Monotone trees deliver the second layer directly. TThe third layer isis appliedied to the slope of the financial transmission function rather than to its level, which makes the function itself convex.

We validate the identified shock against evidence that is not used in estimation. Its largest realizations correspond to the collapse of Lehman Brothers, the COVID-19 market disruption, the acute phase of the euro area sovereign debt crisis, the 9/11 attacks and the WorldCom bankruptcy. It correlates strongly with the monthly change in the composite indicator of systemic stress of \cite{hollo2012ciss} for the US, and the corresponding indicators for the euro area and the UK. Its dispersion is almost twice as large in NBER recessions as in expansions, and its mean is adverse in recessions and slightly benign in expansions. These checks support its interpretation as a financial shock. It is also in line with the evidence that large adverse shocks and economic downturns tend to coincide \citep{tenreyro2016pushing, ascari2022sign}.


Our empirical analysis yields three main results. First, the effects of US financial shocks are strongly asymmetric. For large shocks, an adverse disturbance lowers US industrial production by more than three times as much as a benign disturbance of the same size raises it. Second, the asymmetry grows with the size of the shock. For small shocks the adverse response is only moderately larger than the benign one, and the gap widens steadily as the disturbance becomes larger. On impact, convexity implies that the gap cannot narrow as the shock becomes larger. How quickly it widens, and whether it survives the propagation of the shock, is estimated. A linear VAR cannot generate this pattern. A regime-switching model with linear equations also implies proportional responses when the regime path is held fixed. Third, the asymmetry is broad-based. Prices, interest rates, the excess bond premium and equity prices display asymmetries of a similar size as US output. The same holds for industrial production in the euro area, Japan and Canada, and equity prices fall strongly and immediately in all five economies. Most of the asymmetry thus arises on impact, through the convex transmission function, rather than in the propagation of the shock. UK industrial production is the exception. Its response cannot be distinguished from zero, which reflects how little UK output co-moves with US conditions at monthly frequency.


The remainder of the paper is structured as follows. Section \ref{sec: lp} presents model-free evidence from local projections. Section \ref{sec: econometrics} introduces the econometric framework. Section \ref{sec: empirics} contains the empirical application. It describes the data and the identification scheme, characterizes the identified financial shock and its transmission function, and reports the domestic and international responses to US financial shocks. It also relates the differences across countries to country characteristics. The final section concludes. Additional empirical results are collected in the appendix.

\section{Some empirical evidence}\label{sec: lp}
We start with evidence that does not rely on our model. We identify US financial shocks as in \citet{gilchrist2012ebp}. The shock $\varepsilon_t$ is the innovation in the excess bond premium that is orthogonal to current US industrial production and consumer prices, and to twelve lags of these two series, the excess bond premium, the one-year and ten-year Treasury yields and the returns on the S\&P 500. We standardize it, and positive values are adverse. This recursive scheme rests on timing assumptions that we relax in our model (see Subsection \ref{sec: ident}). Here it serves as a simple and widely used benchmark.

We relate the shock to industrial production in the US, the euro area, the UK, Japan and Canada through local projections \citep{jorda2005estimation},
\begin{equation}
    y_{c,t+h} - y_{c,t-1} = a_{c,h} + \sum_{r=1}^{4} \beta_{r,h} \, \mathbb{I}(\varepsilon_t \in \mathcal{Q}_r) + \bm{g}_{c,h}'\bm{w}_{c,t} + u_{c,t+h}, \quad h = 0, \dots, 24, \label{eq:lp}
\end{equation}
where $y_{c,t}$ is 100 times the log of industrial production in economy $c$ and $\mathbb{I}(\cdot)$ is the indicator function. The sets $\mathcal{Q}_1$ to $\mathcal{Q}_4$ contain the the bottom 5\%, the 5th–20th percentiles, the 80th–95th percentiles, and the top 5\% of shocks (where positive shocks are adverse). The central 60\% of the shocks serve as the reference. The vector $\bm{w}_{c,t}$ contains twelve lags of the growth rates of industrial production in economy $c$ and in the US, of US inflation, of the excess bond premium, of the one-year Treasury yield and of the returns on the S\&P 500.

The coefficient $\beta_{4,h}$ measures the average effect of a shock among the most adverse 5\%, and $\beta_{1,h}$ that of a shock among the most benign 5\%. Both groups are equally likely. The comparison thus mirrors the one we make with our model, in which the shock has a standard Normal distribution (see Subsection \ref{sec: ident}). With symmetric effects, $\beta_{4,h}$ would equal $-\beta_{1,h}$. The tails of the shock distribution are close to symmetric, so both groups also contain shocks of similar size. We estimate Eq.\ (\ref{eq:lp}) for each economy separately, with the standard errors of \citet{newey1987simple}. For the four foreign economies we also estimate it jointly, with country-specific intercepts and controls and common coefficients $\beta_{r,h}$, and use the standard errors of \citet{driscoll1998consistent}. The sample runs from 1974M1 to 2019M12, and from 1992M1 for the euro area. It ends before the COVID-19 period and starts much earlier than the sample of our model. The US series come from FRED-MD, and industrial production abroad from the OECD Main Economic Indicators and the ECB.

\begin{figure}[!htbp]
    \caption{Responses of industrial production to large adverse and large benign US financial shocks. \label{fig:lp}}
    \centering
    \includegraphics[width=\textwidth]{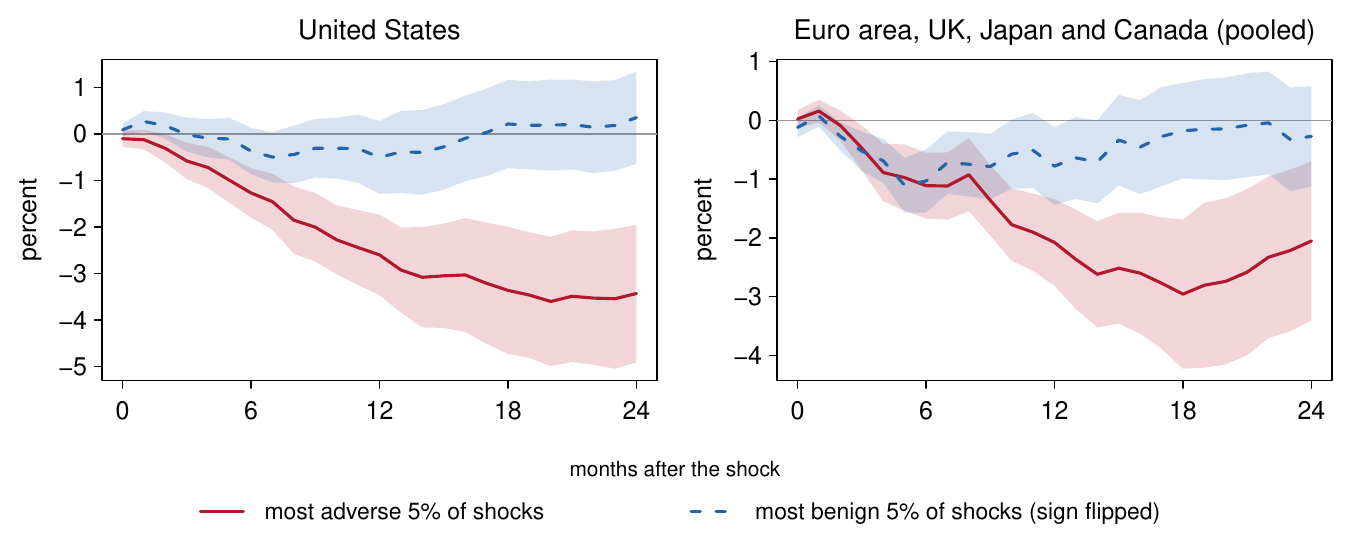}
    \begin{minipage}{\textwidth}
    \vspace{2pt}
    \scriptsize \emph{Note:} Local projections of Eq.\ (\ref{eq:lp}), 1974M1--2019M12. Solid lines show $\beta_{4,h}$, the response to a shock among the most adverse 5\%, and dashed lines show $-\beta_{1,h}$, the sign-flipped response to a shock among the most benign 5\%. Shaded areas are 68\% confidence bands. The right panel pools the euro area, the UK, Japan and Canada.
    \end{minipage}
\end{figure}

Figure \ref{fig:lp} plots $\beta_{4,h}$ and $-\beta_{1,h}$, so that the two lines coincide if the effects are symmetric. In the US, a shock among the most adverse 5\% lowers industrial production steadily over the following one and a half years. The effect of a shock among the most benign 5\% is small throughout and close to zero in the second year. The pooled responses of the four foreign economies show the same pattern. The two responses are similar in the first months. Afterwards, industrial production abroad keeps falling after large adverse US financial shocks, while the small effects of large benign shocks fade. The 68\% bands of the two responses separate within the first year in the US and after about a year abroad. The difference is not estimated precisely enough to reject symmetry at conventional significance levels, since the tails of the shock distribution contain only a few dozen months. Figure \ref{fig:lpcountries} in the appendix reports the four foreign economies separately. Canada shows the same asymmetry, and the euro area does so from about a year onward. In Japan the two responses are close to each other, and for the UK both are small and imprecisely estimated.

These projections have several limitations. First, the tails of the shock distribution contain few observations. The estimates for large shocks are therefore imprecise, and the projections say little about how the effects change with the size of the shock.\footnote{When we allow the effect per unit of the shock to differ between shocks below and above 1.5 standard deviations in absolute value, the effect per unit is, if anything, smaller for large shocks. This measure differs from the one we use with our model. There, the excess bond premium and industrial production inherit the same transmission function. In our estimates, the peak response of US industrial production relative to the impact response of the excess bond premium does not rise with the size of an adverse shock either.} Second, the shock rests on a recursive identification. Timing assumptions of this kind separate financial shocks from other disturbances only imperfectly, since financial and macroeconomic variables interact within the month. Third, the shock is estimated in a first step and then treated as observed, so the uncertainty about it does not enter the confidence bands. Fourth, each projection covers a single variable in a single economy. The projections therefore do not show how prices, interest rates, asset prices and exchange rates respond, and they cannot tell whether the asymmetry arises on impact or through the propagation of the shock. Finally, the sample has to end before the COVID-19 period, since a few observations in 2020 would dominate the estimates.

Our joint modeling framework overcomes these limitations. It models the macroeconomic and financial variables of the five economies jointly, so that the responses of all variables in all economies are consistent with each other. It identifies the financial shock together with four other structural shocks through sign, zero and shape restrictions instead of timing assumptions. Since the shock is estimated together with the rest of the model, the uncertainty about it enters the uncertainty about the responses. The transmission function of the shock is estimated from all its realizations, not only from those in the tails, and the shape restrictions discipline it where the data are scarce. The model thus traces out how the responses depend on the sign and the size of the shock, and it separates the asymmetry on impact from the part added by propagation. Finally, the common volatility scale allows us to include the COVID-19 period without letting it dominate the estimates.

\section{Econometric framework}\label{sec: econometrics}
Our goal is to understand whether financial shocks trigger asymmetric international reactions. This section presents the model and its estimation. Subsection \ref{subsec:model} sets out the observation equation. Subsection \ref{subsec:transmission} introduces the transmission functions through which structural shocks enter it and derives the impact response. Subsection \ref{subsec:bart} describes how we approximate the unknown functions using BART and its monotone version. Subsection \ref{subsec:covid} describes our treatment of the COVID-19 period. Subsection \ref{subsec:priors} presents the priors, Subsection \ref{subsec:mcmc} sketches the MCMC algorithm, and Subsection \ref{subsec:girf} defines the dynamic responses. We discuss identification separately in Subsection \ref{sec: ident}.

\subsection{A nonlinear factor-augmented multi-country model}\label{subsec:model}
Let $\bm y_{it}$ denote an $M_i$-dimensional vector of country-specific macroeconomic and financial time series for countries $i=1, \dots, N$. In our application, $N=5$ and the countries are the US, the euro area, the UK, Japan and Canada. Stacking these variables yields an $M$-dimensional vector $\bm Y_t =(\bm y_{1t}', \dots, \bm y_{Nt}')'$, where $M=\sum_{i=1}^N M_i$. Let $\bm X_t = (\bm Y'_{t-1}, \dots, \bm Y'_{t-P}, 1)'$ collect $P$ lags and an intercept. We decompose $\bm Y_t$ into four additive components,
\begin{equation}
    \bm Y_t = \bm A \bm X_t + \bm \Lambda_\mu \bm \mu(\bm X_t) + \bm \Lambda_q \bm \mu_q(\bm q_t) + \bm e_t, \qquad \bm e_t \sim \mathcal{N}(\bm 0_M, s_t^2 \bm H), \label{eq: obs}
\end{equation}
where $\bm{A}$ is an $M \times (MP+1)$ matrix of linear VAR coefficients, $\bm{\mu}(\bm{X}_t) \in \mathbb{R}^{Q_\mu}$ is a flexible nonlinear common component loaded onto the observables by $\bm{\Lambda}_\mu \in \mathbb{R}^{M \times Q_\mu}$, $\bm{\mu}_q(\bm{q}_t) \in \mathbb{R}^{Q_q}$ is the structurally identified nonlinear block loaded by $\bm{\Lambda}_q \in \mathbb{R}^{M \times Q_q}$, and $\bm H = \operatorname{diag}(H_1, \dots, H_M)$ collects the idiosyncratic variances. The scalar $s_t$ is a common volatility scale. It equals one in normal times, and we describe it in Subsection \ref{subsec:covid}. 

The $Q_\mu$ elements $\mu_j(\bm{X}_t)$ of the first nonlinear block are unknown functions of the lagged observables. These nonlinear factors absorb nonlinear common variation that has no structural interpretation. Examples are state dependence in the propagation mechanism, threshold-like behavior, and general nonlinear relations between $\bm Y_t$ and $\bm X_t$. Note that $\bm A \bm X_t$ and $\bm \Lambda_\mu \bm \mu(\bm X_t)$ are not separately identified since the latter can also capture linear dynamics. Our analysis concerns their sum, $\bm A \bm X_t + \bm \Lambda_\mu \bm \mu(\bm X_t)$, so we do not interpret the two components separately.\footnote{For a more formal discussion on this, see \citet{clark2026flexible}.}

The second nonlinear block is a sequence of $Q_q$ monotone and univariate functions,
\begin{equation*}
    \bm\mu_q(\bm q_t) = \bigl(\mu_{q,1}(q_{1,t}), \dots, \mu_{q,Q_q}(q_{Q_q,t})\bigr)', \qquad \bm q_t \sim \mathcal{N}(\bm 0, \bm I_{Q_q}).
\end{equation*}
The shocks are independent over time and independent of $\bm e_t$. Each function $\mu_{q,j}$ depends on a single latent structural shock $q_{j,t}$ and is non-decreasing in it. We call $\mu_{q,j}$ the transmission function of shock $j$, and we refer to $\mu_{q,j}(q_{j,t})$ as the corresponding structural factor. Subsection \ref{subsec:transmission} explains why we require these functions to be univariate and monotone. The matrix $\bm{\Lambda}_q$ carries additional sign, zero and anchor restrictions described in Subsection \ref{sec: ident} to identify the shocks.

Both $\bm\mu$ and $\bm\mu_q$ are unknown. We approximate them using Bayesian additive regression trees \citep[BART,][]{chipman2010bart}. For $\bm\mu$ we use standard BART, as in \citet{clark2026flexible}. For $\bm\mu_q$ we use monotone BART, which respects the shape restrictions on the transmission functions. Subsection \ref{subsec:bart} introduces both. 

Our model allows for nonlinear impact effects of economic shocks, in line with \citet{bobeica2025beware}. It is also related to a range of other models in the literature. For instance, setting $\bm{\Lambda}_q = \bm{0}$ recovers the specification of \citet{clark2026flexible}. Alternatively, setting $\bm{\Lambda}_\mu = \bm{0}$ and $\mu_{q,j}(q) = q$, while retaining $\bm{\Lambda}_q$, recovers a linear VAR augmented with structural factors.

\subsection{Transmission functions and the impact response}\label{subsec:transmission}
In many of the nonlinear models discussed in the introduction, nonlinearity enters through propagation, while the impact of a shock is linear in its size. Our framework differs in this respect. We place the nonlinearity in a univariate function of the structural shock itself, which we call its transmission function. The impact response then varies with the sign and the size of the shock. The transmission function is univariate, so its shape can be restricted a priori in an economically meaningful way.\footnote{This specification imposes a restriction. A single scalar function governs the impact response of each shock, so the ratio of the adverse to the benign impact response is identical for every variable in the system. The model generates different magnitudes of impact response for benign and adverse shocks, but not different patterns across variables. Any heterogeneity in asymmetry across variables is generated by the linear and BART components through propagation. Richer impact asymmetries would require a multivariate version, which we leave for future work.}

The $\bm \mu_q$ block contains the structurally identified shocks. The impact response of variable $m$ to a disturbance of size $\delta$ in shock $j$, given the realized value $q_{j,t}$ of that shock, equals
\begin{equation}
   \gamma_{mj,0,t}^{(\delta)} = \lambda_{mj} \bigl[ \mu_{q,j}(q_{j,t}+\delta) - \mu_{q,j}(q_{j,t}) \bigr], \label{eq: impact}
\end{equation}
where $\lambda_{mj}$ denotes the $(m,j)$th element of $\bm\Lambda_q$. This is the impact element of the generalized impulse response defined in Subsection \ref{subsec:girf}. The sign and the magnitude of the impact response thus factorize into the loading and the increment of $\mu_{q,j}$ at the realized value of the shock. In a linear model the bracketed term is simply $\delta$ and the impact is proportional to the shock. Here, this is not necessarily the case. The shape of $\mu_{q,j}$ decides whether an adverse shock has a larger impact than a benign one of the same size, and whether the impact increases more than proportionally with shock size. We therefore impose shape restrictions when identifying the shocks. We return to this point in Subsection \ref{sec: ident}. Subsection \ref{subsec:bart} shows how monotone BART builds these restrictions into the approximation of $\mu_{q,j}$.

\subsection{Approximating the unknown functions using BART}\label{subsec:bart}
In principle, we can use any nonlinear learning technique to approximate $\bm\mu$ and $\bm\mu_q$, such as neural networks, Gaussian processes or splines. Because of its empirical success with many different datasets (including datasets commonly employed in macroeconomics and finance) and the existence of a set of hyperparameters that work well in all of these, we approximate the unknown functions through BART \citep{chipman2010bart}. BART has been successfully employed in economics for nowcasting GDP \citep{huber2020nowcasting}, for tail forecasting of output, inflation and unemployment \citep{clark2023tail} and for structural analysis with nonlinear factor models \citep{clark2026flexible}.

BART is a sum-of-trees model. For the $j$th nonlinear factor, it approximates
\begin{equation}
    \mu_j(\bm X_t) \approx \sum_{b=1}^{B} \mathfrak{t}(\bm X_t | \bm\theta_{j,b}, \mathcal{T}_{j,b}), \label{eq: bart}
\end{equation}
where $\mathfrak{t}$ is a tree function, $\bm\theta_{j,b}$ a vector of terminal node parameters, $\mathcal{T}_{j,b}$ a tree structure and $B$ the total number of trees. These tree structures consist of decision rules that take the form $\{X_{k,t} \le c\}$ or $\{X_{k,t} > c\}$, with $X_{k,t}$ denoting the $k$th element of $\bm X_t$ and $c$ a threshold. They therefore partition the input space into disjoint regions. For each of these disjoint sets, there is a terminal node parameter that plays the role of a fitted value in a regression model. Trees are prone to over-fitting if no regularization is introduced. Summing over many (possibly) complex trees further increases the risk of over-fitting. As a solution, BART uses Bayesian regularization priors to force each of the trees to take a particularly simple form and thus act as a weak learner. Summation over many such simple trees produces a model with a great deal of representation flexibility, while regularization to keep the trees simple limits the risk of over-fitting.

We adopt the standard prior specification proposed in \citet{chipman2010bart} with $B = 250$ trees. This prior involves three aspects. First, the probability that a given node is non-terminal decreases in the depth $d$ of the node. It equals $\mathfrak{a}(1+d)^{-\mathfrak{b}}$ with $\mathfrak{a}=0.95$ and $\mathfrak{b}=2$, which favors small trees. Second, the splitting variable assignments are uniformly distributed so that, a priori, each splitting variable is equally likely. Third, conditional on the chosen splitting variable, we use a uniform prior on the discrete set of available splitting values. On the terminal node parameters we use a Gaussian prior. We follow \citet{chipman2010bart} and rescale the data of each BART update, the pseudo-observations defined in Appendix \ref{App:Tec}, to lie between $-0.5$ and $0.5$. Each element of $\bm\theta_{j,b}$ then follows a Gaussian distribution with mean zero and standard deviation $0.5/(\mathfrak{m}\sqrt{B})$, where $\mathfrak{m}=2$. Larger values of $B$ reduce the prior variance, so that each tree acts as a weak learner and the risk of over-fitting is controlled. Since we sum over many trees, each individual tree is not identified. This poses no problem, since our interest centers on the sum over all trees and not on a specific tree.

For the transmission functions we use monotone BART \citep{chipman2022mbart}. Monotone BART is a version of BART in which every tree is constrained to be a monotone function of its arguments. The sum of the trees is then monotone as well. Each transmission function has a single argument, the latent shock $q_{j,t}$. A tree then divides the real line into intervals and assigns a terminal node parameter to each of them. The tree is non-decreasing if these parameters rise from left to right. We therefore write each tree through the jumps between adjacent intervals. Tree $b$ of shock $j$ has split points $c_{j,b,1},\dots,c_{j,b,L_{j,b}}$ and non-negative jumps $\eta_{j,b,1},\dots,\eta_{j,b,L_{j,b}}$. For the four non-financial shocks,
\begin{equation}
    \mu_{q,j}(q) = \alpha_j q + \sum_{b=1}^{B_q} \sum_{l=1}^{L_{j,b}} \eta_{j,b,l} \bigl[ \mathbf{1}(q \geq c_{j,b,l}) - \mathbf{1}(0 \geq c_{j,b,l}) \bigr], \qquad \alpha_j \geq 0, \quad \eta_{j,b,l} \geq 0. \label{eq: mbart}
\end{equation}
This function is a linear term plus a sum of non-decreasing step functions, so it is non-decreasing. For the financial shock we apply the same construction to the slope $\dot\mu_{q,F}(q) = d\mu_{q,F}(q)/dq$ rather than to the level,
\begin{equation}
    \dot\mu_{q,F}(q) = \alpha_F + \sum_{b=1}^{B_q} \sum_{l=1}^{L_{F,b}} \eta_{F,b,l} \mathbf{1}(q \geq c_{F,b,l}), \qquad \alpha_F \geq 0, \quad \eta_{F,b,l} \geq 0, \label{eq: mbartF}
\end{equation}
and integrate from zero. This gives
\begin{equation*}
    \mu_{q,F}(q) = \alpha_F q + \sum_{b=1}^{B_q} \sum_{l=1}^{L_{F,b}} \eta_{F,b,l} \bigl[ (q - c_{F,b,l})_+ - (-c_{F,b,l})_+ \bigr],
\end{equation*}
with $(x)_+ = \max(x, 0)$. The slope is non-negative and non-decreasing, so $\mu_{q,F}$ is non-decreasing and convex. It is piecewise linear, with kinks at the split points. Both constructions satisfy $\mu_{q,j}(0) = 0$, and the level of each function is absorbed by the intercept in $\bm A$. If all trees are empty, the function is linear, $\mu_{q,j}(q) = \alpha_j q$. The linear model is thus nested.

We use $B_q = 20$ trees per function. The split points come from a fixed grid of $81$ quantiles of the standard Normal distribution, the prior distribution of $q_{j,t}$, at equally spaced probabilities from $0.5$ to $99.5$ percent. The grid thus covers the interval $[-2.58, 2.58]$. It is the same in every draw of the sampler and does not depend on the data. Since the grid follows the quantiles, most potential kinks lie where the shocks are located. Beyond the outermost grid points the trees are constant. The non-financial functions are then linear with slope $\alpha_j$, and the financial function is linear with the slope it reaches at the last grid point. For the financial function, this linear continuation is the least convex extension that the shape restriction allows. The number and the location of the steps and kinks are estimated. Unlike a spline basis, the trees therefore do not fix the location of the nonlinearity in advance. Qualitative restrictions on the shape of $\mu_{q,j}$, such as an upward slope or an increasing slope, reduce to non-negativity of $\alpha_j$ and of the jumps. Subsection \ref{sec: ident} states which restrictions we impose and why.

\subsection{The COVID-19 period}\label{subsec:covid}
In the spring of 2020 many of our series moved by an order of magnitude more than in any other month of the sample. In a model with flexible nonlinear components, a few such observations can dominate the estimates. Following \citet{lenza2022vector}, we let the measurement errors share the common scale $s_t$ in Eq.\ (\ref{eq: obs}). We set $s_t = 1$ before March 2020 and treat $s_t$ as a free parameter in March, April and May 2020. From June 2020 onward it decays geometrically to one,
\begin{equation}
    s_t = 1 + (\bar s - 1)\rho^{t - t^*}, \label{eq: covid}
\end{equation}
where $t^*$ denotes May 2020, $\bar s$ is the scale in that month and $\rho \in (0,1)$ is the decay rate. The scale multiplies the errors of all variables at once. It does not affect the structural shocks $\bm q_t$, which keep their standard Normal distribution. In effect, the months of the pandemic receive less weight in the estimation of all other parameters, and the amount of down-weighting is estimated.

\subsection{Priors}\label{subsec:priors}
We use the priors proposed in \citet{clark2026flexible} for the linear and BART blocks. The VAR coefficients receive an element-wise horseshoe prior around a Minnesota random walk prior mean. Writing $A^*_{mk}$ for the deviation of the $(m,k)$th coefficient from that mean, we set $A^*_{mk} | \psi_{mk} \sim \mathcal{N}(0, \psi_{mk})$ with $\psi_{mk} = \nu_{mk}^2\tau^2$ and $\nu_{mk}, \tau \sim C^+(0,1)$. The BART loadings $\bm\Lambda_\mu$ receive a horseshoe prior with a separate global shrinkage parameter for each row, that is, for each equation. In addition, the prior variances of the loadings on the $j$th BART factor are multiplied by $\varpi_j = \varpi/j^{\kappa}$, with $\kappa = 2$ and an estimated scale $\varpi$. This pushes higher-indexed factors toward zero, so that $Q_\mu$ acts as an upper bound and the shrinkage prior selects the effective number of nonlinear factors. The BART component uses the prior described in Subsection \ref{subsec:bart}. The idiosyncratic variances follow weakly informative inverse-gamma priors so that $H_m \sim \mathcal{IG}(0.01, 0.01)$.

Let $\bm\lambda_m$ denote the $m$th row of $\bm\Lambda_q$, written as a column vector, so that $\lambda_{mj}$ is its $j$th element. The loadings on the structural factors feature equation-specific Normal priors truncated to a restriction set $\mathcal{R}_m~(m=1, \dots, M)$:
\begin{equation*}
    \bm\lambda_m | \mathcal{R}_m \sim \mathcal{N}_{\mathcal{R}_m}\bigl(\bm 0, \bm\Sigma_{\Lambda,m}\bigr),
\end{equation*}
where $\bm\Sigma_{\Lambda,m}$ is a diagonal matrix of prior variances.  The diagonal elements of this matrix are set as follows. Entries subject to a sign or anchor restriction receive a relatively large prior variance of $10^2$, since the truncation shrinks the parameter space. For the unrestricted entries of the financial column, which mostly belong to foreign variables, we tighten the prior variance to $8$. The prior variances in the remaining columns are kept at $10^2$.

The trees of the transmission functions use the same tree prior as the BART component, with split points drawn uniformly from the grid points available in a node. The jumps receive half-Normal priors, $\eta_{j,b,l} \sim \mathcal{N}_{+}(0, \omega_j^2)$, and the linear coefficients receive $\alpha_j \sim \mathcal{N}_{+}(0,1)$. We calibrate the scale $\omega_j$ as follows. In prior expectation, the trees raise the slope of the financial transmission function by $0.08$ per unit of $q$, so they add curvature to it. For the other four functions, they raise the level by $0.4$ per unit of $q$. For the four non-financial functions the scale is estimated, with an inverse-gamma prior on $\omega_j^2$ centered on its calibrated value.

For the financial transmission function we additionally require its slope at zero to be at least $0.25$, so that $\dot\mu_{q,F}(0) \geq 0.25$. Since the excess bond premium is anchored at one, this bounds the impact response of the credit spread to a small shock from below. We view the bound as a relevance condition. A disturbance that leaves the credit spread almost unchanged is hard to interpret as a financial shock, even if it moves equity prices. The bound plays a similar role as the bounds on impact elasticities in \citet{kilian2012why}. The joint prior is the prior described above restricted to this event, and the jump scale of the financial function stays at its calibrated value. Without the bound, the financial shock is less clearly separated from the other structural shocks. Its correlation with one of them then becomes sizable.

The scales of March, April and May 2020 receive Pareto priors with scale and shape equal to one, so that $s_t \geq 1$. The decay rate $\rho$ receives a Beta prior with mean $0.8$ and standard deviation $0.2$.

\subsection{Posterior simulation}\label{subsec:mcmc}
We estimate the posterior using Markov chain Monte Carlo (MCMC), combining Gibbs updates with Metropolis--Hastings (MH) steps. Each sweep draws $\bm A$ equation by equation and $\bm\Lambda_\mu$ row by row from their conditional Normal distributions, then updates the BART components by Bayesian backfitting \citep{chipman2010bart}. We draw the structural loadings $\bm\Lambda_q$ row by row from truncated Normal distributions and update $\bm q_t$ for each $t$ using a random walk MH proposal with Robbins--Monro adaptation. The trees of the transmission functions are updated one at a time. For each tree we propose to add, remove or move a split point. The jump at that point is integrated out in closed form, and all jumps of the tree are then drawn from truncated Normal distributions. The sweep ends with updates of the COVID-19 scale, the idiosyncratic variances and the shrinkage hyperparameters. Appendix \ref{App:Tec} describes these updates. We run three independent chains of $20{,}000$ draws each and discard the first $10{,}000$ draws of each chain as burn-in. Of the remaining draws we keep every tenth, which leaves $3{,}000$ draws in total. All results below are based on these draws.\footnote{We ran four further chains with different seeds. In two of them the common volatility scale settles at a markedly lower level in the spring of 2020. Pooling all seven chains leaves the impulse responses almost unchanged.} Table \ref{tab:conv} in the appendix reports convergence diagnostics.

\subsection{Nonlinear dynamic responses}\label{subsec:girf}
Because the model is nonlinear, we use generalized impulse response functions \citep[GIRFs, see][]{koop1996impulse} to compute dynamic causal effects. The response at date $t$ to a disturbance of size $\delta$ in shock $j$ is
\begin{equation}
    \bm{\gamma}_{j,h,t}^{(\delta)} = \mathbb{E}(\bm{Y}_{t+h} | q_{j,t}+\delta, \bullet) - \mathbb{E}(\bm{Y}_{t+h} | q_{j,t}, \bullet),\label{eq:GIRF}
\end{equation}
which is the difference between the expected values of two predictive distributions. The first conditions on the realized shock being perturbed by $\delta$, the second on the shock as realized, and $\bullet$ collects the remaining shocks at date $t$ and the history of the system. We denote the $m$th element of $\bm{\gamma}_{j,h,t}^{(\delta)}$ by $\gamma_{mj,h,t}^{(\delta)}$, and Eq.\ (\ref{eq: impact}) gives its value at $h=0$.

We compute both expectations by iterating the model forward from date $t$ through the VAR component, the BART functions and the transmission functions, with all future shocks and measurement errors set to zero. With linear propagation this gives the GIRF exactly, since future shocks enter both paths in the same way and cancel. With the BART component it is an approximation. The responses depend on the date $t$ through the history of the system and through the realized shock $q_{j,t}$. We average them over the sample. For each retained draw, we pick a date $t$ at random from the estimation sample and compute $\bm{\gamma}_{j,h,t}^{(\delta)}$ at that date. Across draws, this averages the responses over the parameters and over the $T$ dates of the sample. The posterior distributions we report therefore reflect both parameter uncertainty and the variation of the responses across histories. We compute responses on a grid $\delta \in \{-5,-4,\dots,-1,+1,\dots,+5\}$ over a $60$-month horizon. We report the first $24$ months, beyond which the profiles are increasingly dominated by mean reversion.

As a linear benchmark we use a linear version of our model. It sets $\mu_{q,j}(q) = \alpha_j q$ for all five shocks and $\bm\Lambda_\mu = \bm 0$. This is a linear VAR augmented with structural factors (see Subsection \ref{subsec:model}). We estimate it on the same data, with the restrictions in Table \ref{tab: signrest}, the same bound on the slope of the financial transmission function at zero, and the same COVID-19 scale. In both models the structural shocks have unit variance and the loading of the excess bond premium equals one. A shock of size $\delta$ therefore has the same meaning in both models, and we compare their responses without any rescaling. Since the identification is the same, differences between the two models reflect the nonlinear components of our model.

\section{Dynamic responses to US-based financial shocks}\label{sec: empirics}
This section presents the empirical application. We first describe the data and the model specification in Subsection \ref{sec: data}. Subsection \ref{sec: ident} then develops the identification scheme, and Subsection \ref{sec: shock} characterizes the identified financial shock and its transmission function. Subsections \ref{sec: res_domestic} and \ref{sec: res_spillovers} report the domestic and the international responses to US financial shocks, and Subsection \ref{sec: xreg} relates the differences across countries to country characteristics.

\subsection{Data and model specification}\label{sec: data}

Our dataset is monthly and runs from 1999M1 to 2023M9. It contains macroeconomic and financial time series that are standard in multi-country models \citep{scholl2008new, feldkircher2016international, georgiadis2017bi, bai2022macroeconomic}. We estimate the model jointly for five economies, the US, the euro area, the UK, Japan and Canada.

The model includes $M = 31$ series. For the US we use industrial production, the CPI, the shadow rate of \citet{krippner2013ssr}, the investment-to-GDP ratio, the S\&P 500, the excess bond premium (EBP) of \citet{gilchrist2012ebp} and the 10-year government bond yield. The restrictions in Table \ref{tab: signrest} are imposed on the first six of these. The investment ratio and the equity index allow us to separate investment shocks from financial shocks, as in \citet{furlanetto2019identification}. For each of the four foreign economies we use industrial production, the HICP or CPI, the shadow rate, the bilateral exchange rate against the US dollar, a headline equity index and the 10-year government bond yield. The equity indices are the EURO STOXX 50 for the euro area, the FTSE 100 for the UK, the share price index of the Tokyo Stock Exchange for Japan and the S\&P/TSX Composite for Canada.

The series come from several sources. FRED and its monthly database FRED-MD provide US industrial production, US CPI, UK CPI, the USD/EUR and USD/GBP exchange rates, the 10-year government bond yields of the US, the euro area and the UK, the S\&P 500, and US investment and GDP for the investment ratio. The Statistical Data Warehouse (SDW) of the ECB provides EA industrial production, HICP, the EURO STOXX 50 and the composite indicator of systemic stress \citep{hollo2012ciss}. We use the latter only for validation in Subsection \ref{sec: shock}. UK industrial production comes from the Office for National Statistics (ONS), and the FTSE 100 from Macrobond. Shadow rates are taken from \citet{krippner2013ssr} and the excess bond premium from \citet{gilchrist2012ebp}. For Japan and Canada, industrial production, the 10-year government bond yields and the share price indices are taken from the OECD Main Economic Indicators, which we obtain from FRED. Consumer prices come from the BIS and the exchange rates against the US dollar from FRED-MD. US investment and GDP are available only at quarterly frequency. We interpolate them to monthly frequency with the EM algorithm of \citet{stock2002macroeconomic}, so the monthly values of the investment ratio also draw on information from the surrounding quarters. Appendix \ref{App:Data} lists all series with their transformation codes.

Most series enter in log-differences (activity, prices, equity indices, exchange rates) or in first differences (shadow rates, government bond yields). The excess bond premium and the investment ratio enter in levels. We cumulate the impulse responses of the differenced variables to obtain level responses. All variables are standardized before estimation, and all responses are mapped back to original units. Responses reported in percent are 100 times log changes. Interest rate responses are in basis points, and EBP responses are in percentage points.

We use $P=12$ lags, so the estimation sample runs from 2000M1 to 2023M9 and contains $T=285$ observations. We also use $Q_\mu = 10$ BART factors and $Q_q = 5$ structural factors. Each transmission function uses $B_q = 20$ trees on a grid of $81$ split points. The linear benchmark of Subsection \ref{subsec:girf} uses the same lags and the same MCMC settings. In both models, the slope of the financial transmission function at zero sits at its lower bound. The linear benchmark thus nearly coincides with the tangent of the nonlinear transmission function at zero, and the two models agree on the impact of small shocks. Appendix \ref{App:linbart} reports a second linear benchmark, which keeps the BART component of our model.

\begin{figure}[!htbp]
    \caption{Common volatility scale of the measurement errors around the COVID-19 period. \label{fig:covid}}
    \centering
    \includegraphics[width=0.75\textwidth]{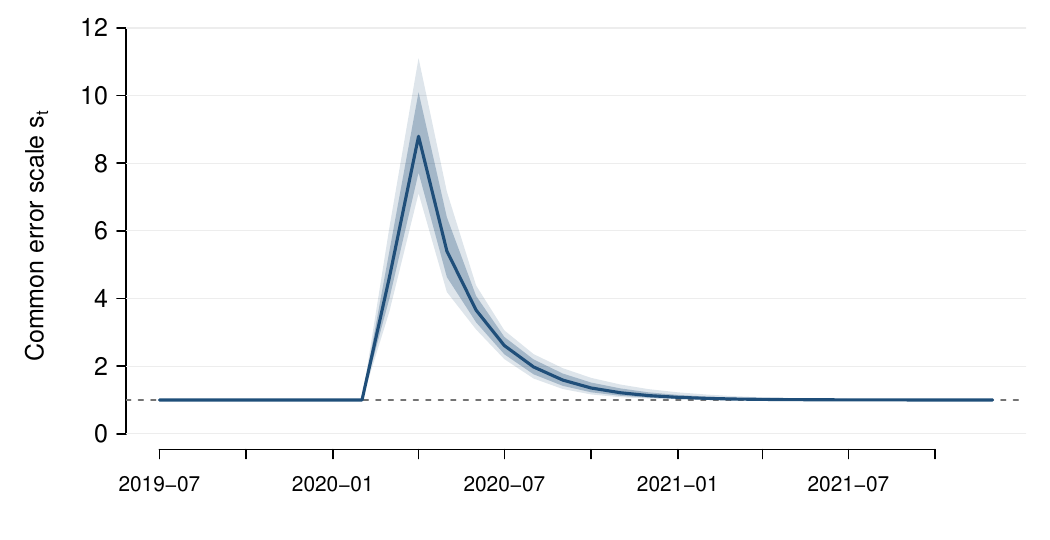}
    \begin{minipage}{\textwidth}
    \vspace{2pt}
    \scriptsize \emph{Note:} Posterior median of $s_t$ in Eq.\ (\ref{eq: covid}) with 16--84\% and 5--95\% credible bands. The scale equals one before March 2020 by construction. The dashed line marks one.
    \end{minipage}
\end{figure}

Figure \ref{fig:covid} shows the posterior of the common scale. It jumps in March 2020 and peaks in April, when the measurement errors are about an order of magnitude more volatile than in normal times. It then declines steadily and is close to one by the spring of 2021.

\subsection{Identifying financial shocks when transmission is nonlinear}\label{sec: ident}
We identify the structural shocks through restrictions on the loadings $\bm\Lambda_q$. In a linear structural VAR, a sign restriction on an impact coefficient is the same as a sign restriction on the impact response. In our model this is not automatically the case. By Eq.\ (\ref{eq: impact}), the impact response is the product of the loading and the increment of $\mu_{q,j}$ at the realized value of the shock. If $\mu_{q,j}$ sloped downward for some values of the shock, a positive disturbance would there produce a response with the opposite sign to the loading. Since every transmission function is non-decreasing, its increment is non-negative for every value of the shock and every $\delta>0$. A nonzero impact response thus has the sign of its loading. Monotonicity therefore turns sign restrictions on the loadings into sign restrictions on the impact responses, as in a linear model. Monotone BART delivers it directly, so no rejection step over rotation matrices is needed.\footnote{Related uses of monotonicity constraints to impose economically motivated structure in tree-based inflation models include \citet{buckmann2025blockwise}, although their approach does not identify structural shocks directly.}

Following \citet{furlanetto2019identification}, we identify five structural disturbances from the impact loadings on six US variables. These are supply ($S$), demand ($D$), monetary policy ($\mathit{MP}$), investment ($I$) and financial ($F$) shocks, and we index them accordingly, $j \in \{S, D, \mathit{MP}, I, F\}$. Table \ref{tab: signrest} reports the restrictions. Two of them deserve a closer look. The first separates the financial shock from the investment shock. A favorable shock to the marginal efficiency of investment raises investment but lowers the relative price of installed capital. A favorable financial shock raises both. The second separates the financial shock from monetary policy. A contractionary financial shock lowers the policy rate and raises the credit spread, whereas a contractionary policy shock raises the policy rate. This addresses the concern of \citet{caldara2016macroeconomic} and \citet{barnichon2022effects} that identified financial shocks are often contaminated by policy. The financial column is oriented so that a positive realization is contractionary.

In addition, we fix one loading per shock, which we call an anchor. Anchors appear in bold in Table \ref{tab: signrest}. They normalize the scale shared by each loading column and its transmission function, since rescaling a column of $\bm\Lambda_q$ and inversely rescaling its transmission function leaves the likelihood unchanged. For the financial shock, the excess bond premium loading is anchored at $+1$. The other loadings then measure impact responses relative to the excess bond premium, in standardized units.

\begin{table}[!htbp]
\caption{Sign, zero and anchor restrictions on the impact loadings $\bm\Lambda_q$. \label{tab: signrest}}
\centering \small
\begin{tabular}{lccccc}
\toprule
 & Supply & Demand & Monetary & Investment & Financial \\
\midrule
US industrial production & $+$      & $\bm{+1}$ & $-$      & $+$ & $-$ \\
US CPI                   & $\bm{-1}$& $+$       & $-$      & $+$ & $-$ \\
US shadow rate           & $\cdot$  & $+$       & $\bm{+1}$& $+$ & $-$ \\
US investment ratio      & $\cdot$  & $-$       & $\cdot$  & $\bm{+1}$ & $-$ \\
US S\&P 500              & $+$      & $\cdot$   & $\cdot$  & $-$ & $-$ \\
US excess bond premium   & $\cdot$  & $\cdot$   & $\cdot$  & $\cdot$ & $\bm{+1}$ \\
\midrule
US 10-year government yield & $\cdot$ & $\cdot$ & $\cdot$ & $\cdot$ & $\cdot$ \\
Foreign industrial production & $\cdot$ & $\cdot$ & $\cdot$ & $\cdot$ & $\bm{0}$ \\
Other foreign variables & $\cdot$ & $\cdot$ & $\cdot$ & $\cdot$ & $\cdot$ \\
\bottomrule
\end{tabular}
\begin{minipage}{\textwidth}
\vspace{4pt}
\scriptsize \emph{Note:} $+$ and $-$ denote sign restrictions on the impact loading, and $\cdot$ denotes an unrestricted entry. Bold entries are point restrictions. The anchors ($\pm 1$) are imposed as tight intervals around the stated value in standardized units, and the zero restrictions hold exactly. The financial shock is oriented so that a positive realization is contractionary, which is why its column carries the signs of an adverse shock. The impact responses of foreign industrial production to the financial shock are restricted to zero. All other foreign entries are unrestricted.
\end{minipage}
\end{table}

All sign restrictions involve US variables only, and the anchor is placed on the US excess bond premium. This is what makes $q_{F,t}$ a US financial shock. We also restrict the impact response of industrial production in the four foreign economies to zero. The shock originates in US financial markets, and it reaches production abroad through trade and financial linkages, which takes time. Asset prices, interest rates, exchange rates and consumer prices can react within the month, so we leave all other foreign loadings unrestricted. The restriction is also in line with the US evidence. The loading of US industrial production, which is only restricted in sign, is close to zero (see Subsection \ref{sec: shock}).\footnote{Without the restriction, the loading of UK industrial production would rest on a few months with large idiosyncratic movements in UK production, such as the rebound after the extra bank holiday of June 2002.}

Finally, we require the financial transmission function to be convex. For any value $q$ of the shock and any $\delta>0$, convexity implies
\begin{equation}
    \mu_{q,F}(q+\delta) - \mu_{q,F}(q) \geq \mu_{q,F}(q) - \mu_{q,F}(q-\delta). \label{eq: asym}
\end{equation}
Both sides are non-negative, since the function is non-decreasing. Combined with Eq.\ (\ref{eq: impact}), the absolute impact response to an adverse shock is at least as large as that to an equally sized benign shock of the same size, for every value of the shock and every variable. This restriction is familiar from the credit shock literature. Adverse credit supply disturbances tighten borrowing constraints and force deleveraging, while favorable ones relax constraints that many borrowers did not face \citep{gilchrist2012ebp, he2013intermediary, christiano2014risk}. Contractionary financial shocks are also found to matter much more than expansionary ones \citep{barnichon2022effects, forni2024nonlinear}. Convexity also implies that the impact per unit of shock does not fall with $\delta$ for adverse shocks and does not rise with $\delta$ for benign shocks. On impact, the gap between the two can therefore only widen as the shock becomes larger. Convexity thus fixes the direction of the impact asymmetry and the direction in which it changes with the size of the shock. Its size, and its evolution over subsequent horizons, are estimated. The linear case remains admissible. The other four transmission functions are non-decreasing, but their curvature is unrestricted.

The shape of $\mu_{q,F}$ is defined relative to the distribution of the shock. Since $q_{F,t}$ is latent, any monotone transformation of it can be absorbed by the transmission function. The standard Normal distribution of $q_{F,t}$ serves as the normalization. An adverse and a benign shock of the same size are therefore equally likely, and our measure of asymmetry compares disturbances of equal probability. A skewed distribution of financial disturbances with linear transmission would look the same in our framework as a convex transmission of Normal shocks, and we do not attempt to separate the two. The shock is also homoskedastic outside the COVID-19 period. Large, adverse and recessionary realizations tend to coincide (see Subsection \ref{sec: shock}), so part of the estimated asymmetry may reflect a higher volatility of financial shocks in downturns.

\subsection{The identified financial shock and its transmission function}\label{sec: shock}
Before examining dynamic responses, we describe the timing and sign of the identified financial shock, compare it with indicators not used in estimation, and examine its estimated transmission function.

\begin{figure}[!htbp]
    \caption{The structurally identified financial shock over time. \label{fig:shockseries}}
    \centering
    \includegraphics[width=\textwidth]{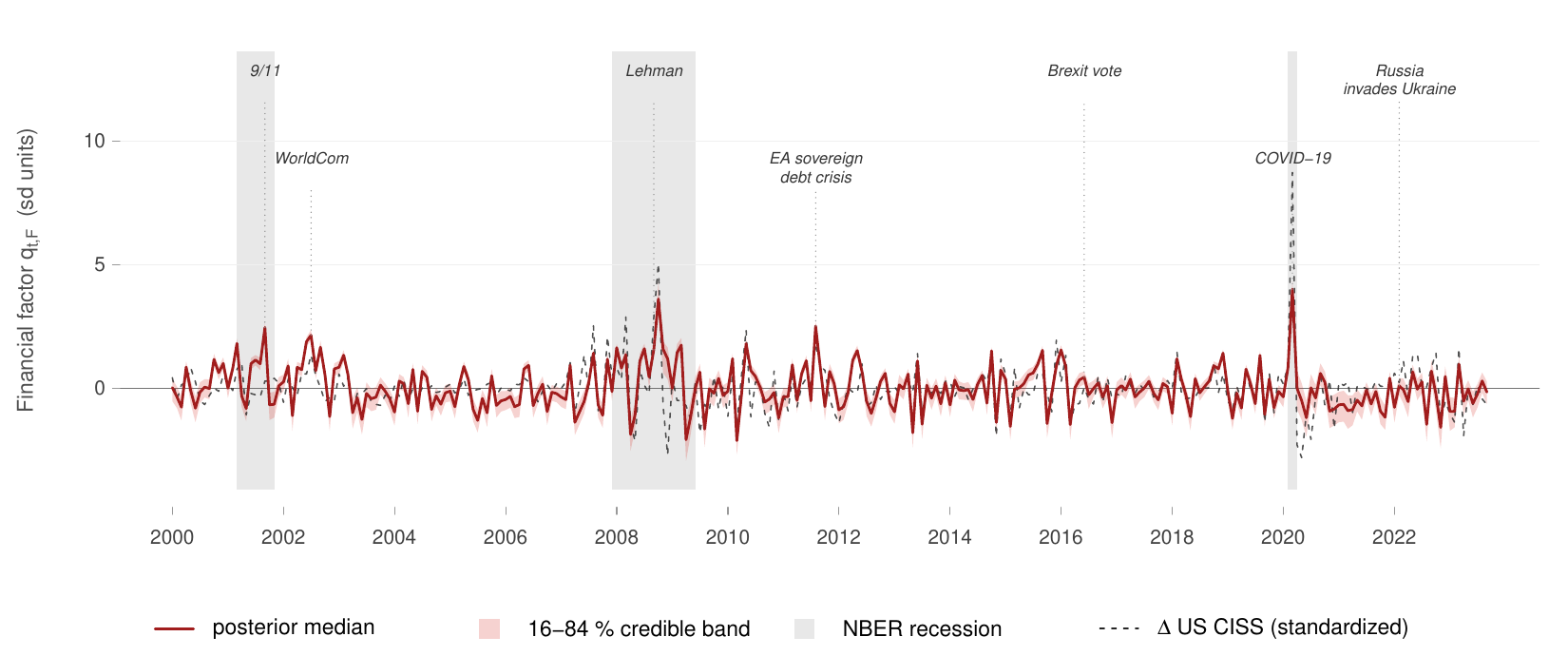}
    \begin{minipage}{\textwidth}
    \vspace{2pt}
    \scriptsize \emph{Note:} Posterior median of $q_{F,t}$ with 16--84\% credible band. Positive values are contractionary. The dashed line is the monthly change in the US composite indicator of systemic stress \citep{hollo2012ciss}, standardized over the estimation sample. Shaded areas are NBER-dated US recessions. Dotted lines mark selected episodes.
    \end{minipage}
\end{figure}

Figure \ref{fig:shockseries} plots the identified financial shock $q_{F,t}$ over time. Its largest realizations line up with well-known episodes of financial stress. The two largest are the COVID-19 market disruption in March 2020 and October 2008, the month after the collapse of Lehman Brothers. They are followed by the euro area sovereign debt crisis in August 2011, the 9/11 attacks in September 2001 and the WorldCom bankruptcy in July 2002.

Figure \ref{fig:shockseries} also compares the shock with a prominent measure of financial conditions. The dashed line is the monthly change in the US composite indicator of systemic stress (CISS) of \citet{hollo2012ciss}. The two series move closely together, and their largest spikes coincide in late 2008 and in March 2020. The shock is about as strongly correlated with the corresponding indicators for the euro area and the UK. The CISS also draws on equity market data, so part of this co-movement reflects the strong response of equity prices to the shock (see below). The shock is also close to orthogonal to the four remaining structural shocks. Finally, the shock is almost twice as dispersed in NBER recessions as in expansions, and adverse realizations are more common in recessions.

\begin{figure}[!htbp]
    \caption{Posterior mean loadings on the structural factors. \label{fig:lambdaq}}
    \centering
    \includegraphics[width=\textwidth]{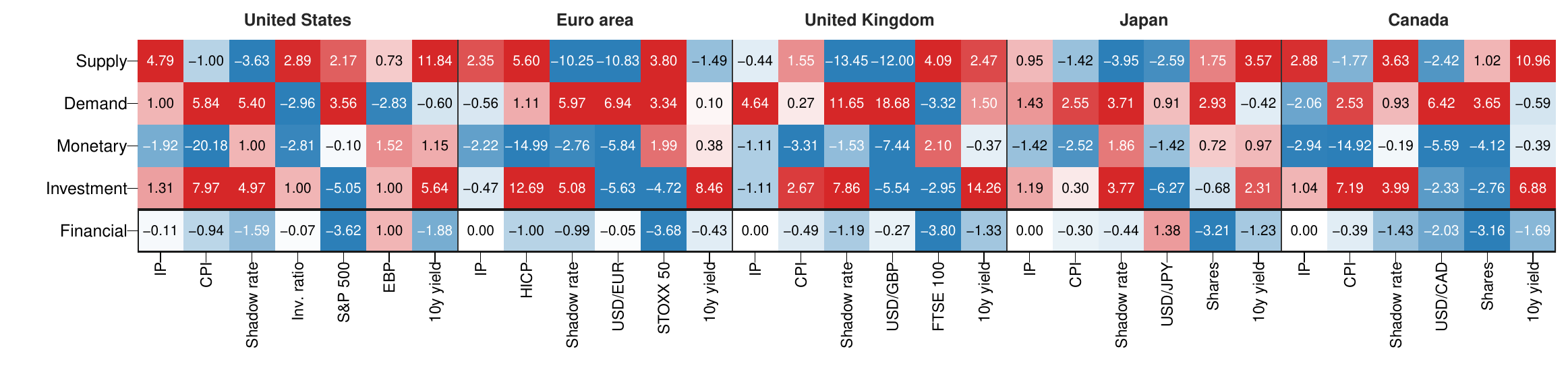}
    \begin{minipage}{\textwidth}
    \vspace{2pt}
    \scriptsize \emph{Note:} Posterior mean of $\bm\Lambda_q$ in standardized units, with colors limited to $[-3,3]$ for illustration. Rows are the structural factors and columns the variables, grouped by country. The frame marks the financial factor. Its entry for the US excess bond premium is the anchor, and its entries for foreign industrial production are restricted to zero. All other foreign entries are estimated. The loadings of the four non-financial factors are large in some places because their transmission functions are flat for most realizations.
    \end{minipage}
\end{figure}

Figure \ref{fig:lambdaq} reports the posterior mean of the impact loadings $\bm\Lambda_q$. We focus on the bottom row, which corresponds to the financial factor. All entries are in standardized units. The excess bond premium is the anchor and loads with $+1$, so every other entry measures the impact response of a variable relative to that of the credit spread. For the US, Table \ref{tab: signrest} restricts the signs of five further entries in this row, but not their size. Equity prices have the largest loadings on the financial factor. In absolute value they are more than three times as large as the anchor, and they are similar across the five equity indices. Interest rates load negatively in all five economies, so that both shadow rates and long-term yields decline on impact. The loadings of US industrial production and of the US investment ratio are close to zero, and those of foreign industrial production are zero by assumption. Consumer prices also load negatively everywhere, but less strongly than equity prices. The loadings of the euro and sterling exchange rates are close to zero. The yen loads positively and the Canadian dollar negatively. On impact, an adverse shock thus lets the yen appreciate and the Canadian dollar depreciate against the US dollar.

\begin{figure}[!htbp]
    \caption{Contribution of the identified financial factor to the observables. \label{fig:hd}}
    \centering
    \includegraphics[width=\textwidth]{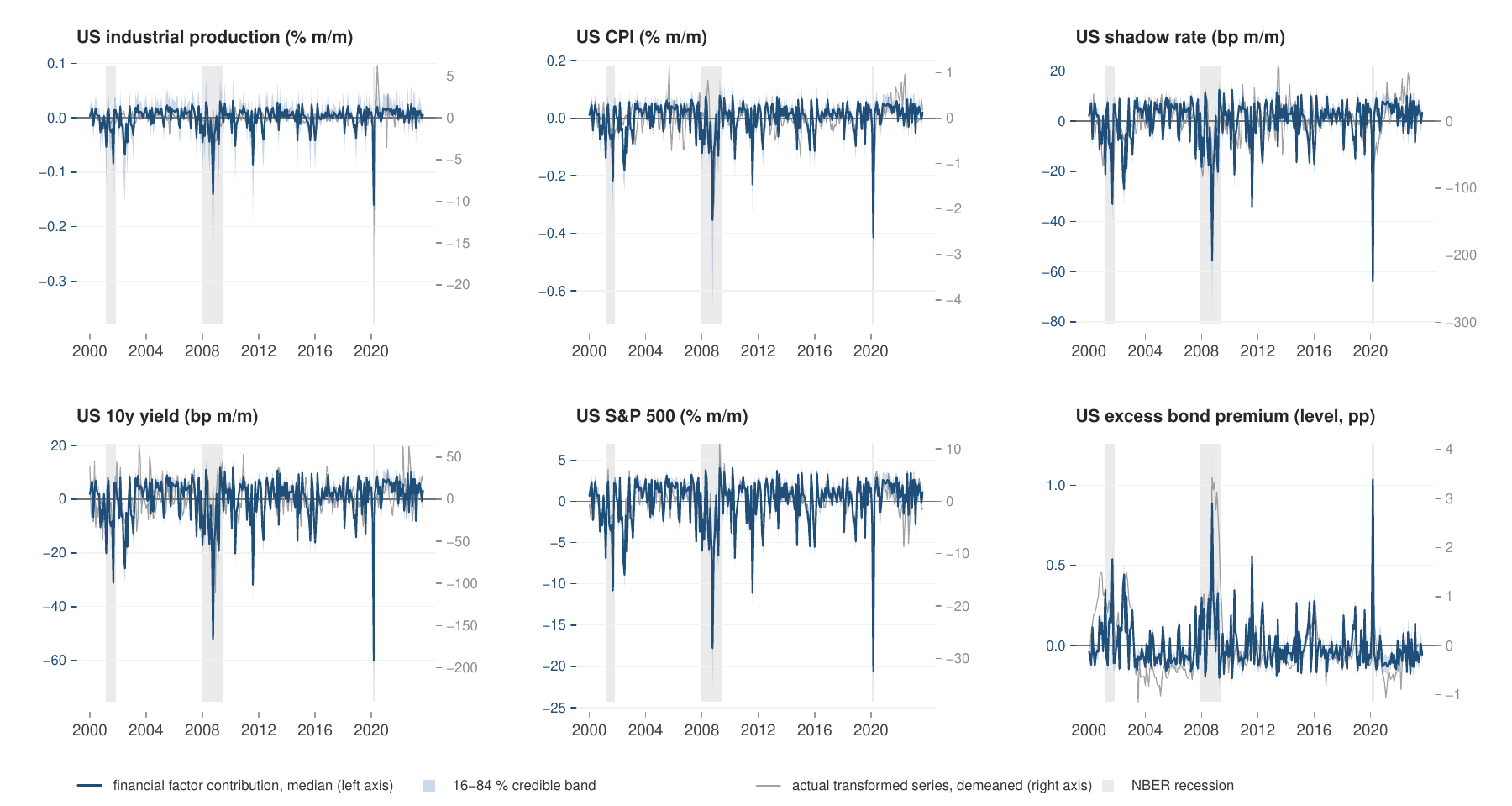}
    \begin{minipage}{\textwidth}
    \vspace{2pt}
    \scriptsize \emph{Note:} Contemporaneous contribution of the financial factor, $\lambda_{mF}\mu_{q,F}(q_{F,t})$, as a deviation from its own sample mean and not cumulated. Posterior median with 16--84\% credible band (left axis). The grey line is the actual series, demeaned (right axis). The two axes share the zero line but differ in scale. Units are the ones in which each series enters the model, that is, monthly percent changes for industrial production, CPI and the S\&P 500, monthly changes in basis points for the shadow rate and the 10-year yield, and percentage points for the level of the excess bond premium. Shaded areas are NBER-dated US recessions.
    \end{minipage}
\end{figure}

Figure \ref{fig:hd} shows the financial factor's contribution to the US variables. It shows the contemporaneous contribution $\lambda_{mF}\mu_{q,F}(q_{F,t})$ for a selection of variables, in the units in which each variable enters the model and as a deviation from its own mean. These are monthly growth rates or monthly changes for all variables except the excess bond premium, which enters in levels. The grey line shows the actual series in the same units. Each panel shows the contemporaneous contribution of the same financial factor, scaled by the variable's loading in Figure \ref{fig:lambdaq}. Up to sign and scale, all contributions thus follow the same path. 

The financial factor's contribution to the S\&P 500 closely tracks the timing of actual returns. It accounts for most of the large monthly falls in equity prices, such as those in October 2008, August 2011 and March 2020. For the shadow rate and the 10-year yield the contribution is sizable as well. In March 2020 it explains more than half of the fall in the shadow rate and most of the fall in the 10-year yield. The excess bond premium is a special case. The contribution captures its sharp increases in October 2008 and March 2020. But the actual premium stays high until the spring of 2009, while the contribution falls back after a single month. The contemporaneous financial contribution alone therefore does not explain this persistence. For consumer prices the contribution is small compared with the actual movements, and for industrial production it is negligible. Even the collapse of output in the spring of 2020 is not attributed to the financial factor within the month. Figures \ref{fig:hdea} to \ref{fig:hdca} in the appendix show the same decomposition for the euro area, the UK, Japan and Canada. The contributions to foreign equity prices track the actual series as closely as in the US. Foreign industrial production receives no contribution within the month, since its loading on the financial factor is restricted to zero. Much of the effect on real activity develops over subsequent months, as shown in the following subsections. The figure also shows that the global financial crisis and the COVID-19 episode unfolded quite differently. During the financial crisis, the factor showed a sequence of mostly adverse realizations from mid-2007 to March 2009, with the largest one in October 2008. The COVID-19 episode consists of a single large adverse month in March 2020, followed by mostly small and often benign realizations.

\begin{figure}[!htbp]
    \caption{Posterior of the financial transmission function $\mu_{q,F}$. \label{fig:muq}}
    \centering
    \includegraphics[width=0.6\textwidth]{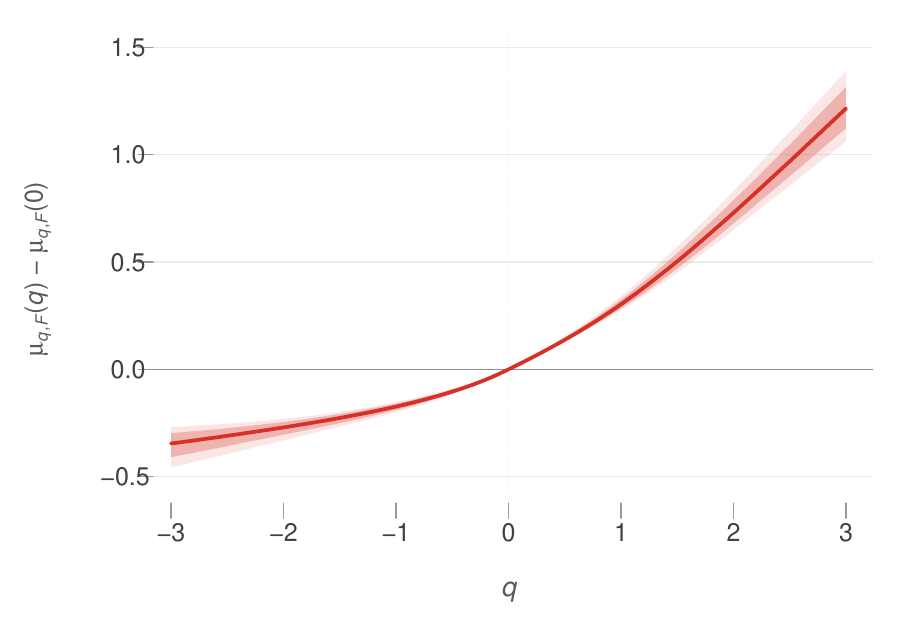}
    \begin{minipage}{\textwidth}
    \vspace{2pt}
    \scriptsize \emph{Note:} Posterior median (solid) with 16--84\% and 5--95\% credible bands, recentered so that every draw passes through the origin. Positive values of $q$ are contractionary. The function is constrained to be non-decreasing and convex.
    \end{minipage}
\end{figure}

Figure \ref{fig:muq} plots the posterior of the financial transmission function. It is non-decreasing and convex by construction, but the degree of curvature is estimated. The function is far from linear. Its slope is small for benign realizations, sits at its lower bound at zero and rises steadily for adverse realizations. For large shocks, an adverse shock thus has roughly four times the impact of a benign shock of the same size. Since the loading cancels from this ratio, every variable inherits this asymmetry on impact. Beyond the outermost grid point the function is linear by construction (see Subsection \ref{subsec:bart}), and only a few realizations of the shock lie in this range. The responses to shocks of four and five standard deviations therefore extend the estimated curvature beyond the bulk of the data. Since the linear continuation is the least convex extension of the function, this tends to understate rather than overstate the asymmetry of large shocks.

\subsection{Asymmetries in domestic reactions of the US economy}\label{sec: res_domestic}

This subsection reports the domestic US responses. We first document the asymmetry in real activity and how it changes with the size of the shock. We then show that asset prices display an asymmetry of a similar size. Finally, we compare both with the linear version of the model.

\begin{figure}[!htbp]
    \caption{Peak responses of US variables to US financial shocks. \label{fig:box_US}}
    \centering
    \includegraphics[width=\textwidth]{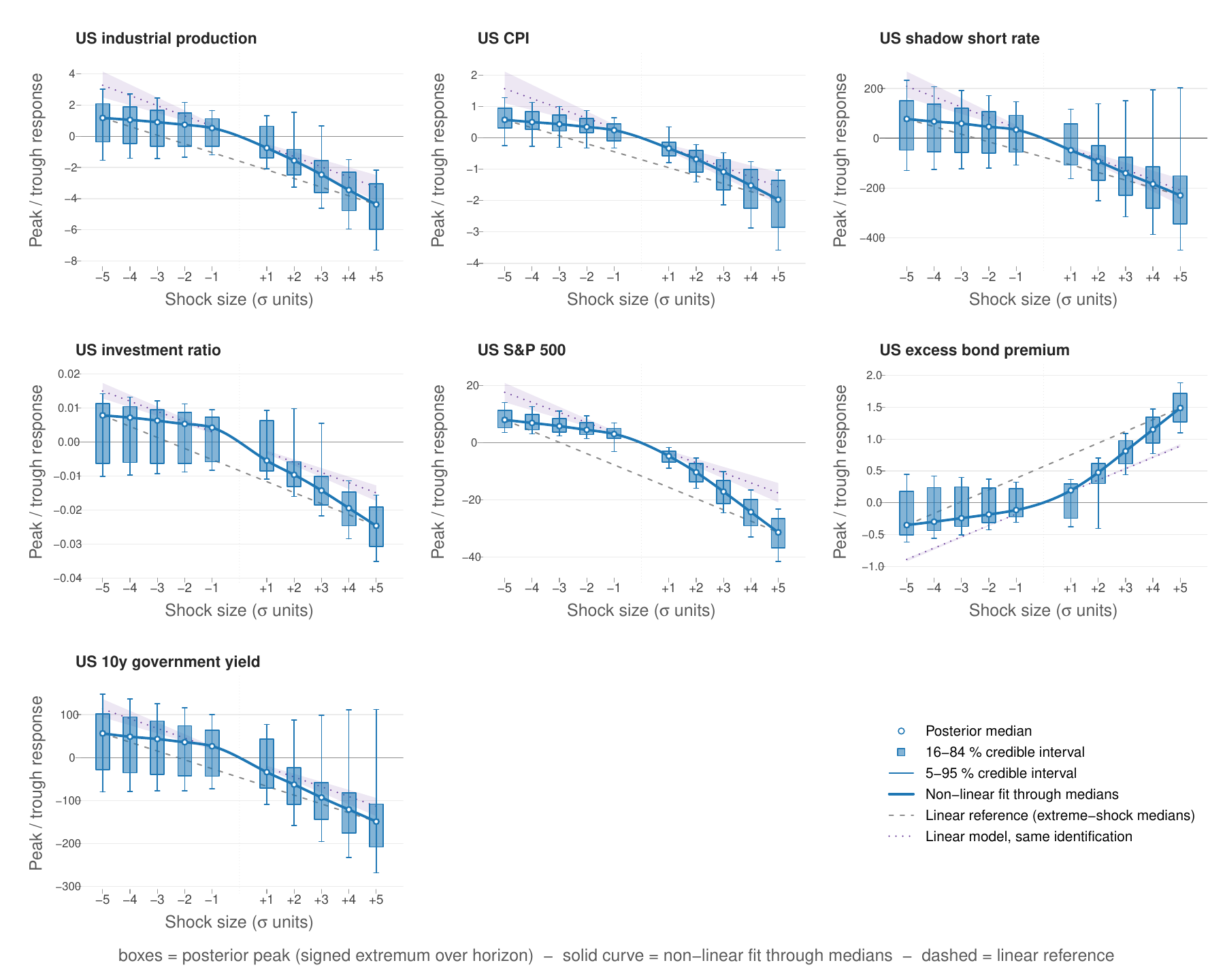}
    \begin{minipage}{\textwidth}
    \vspace{2pt}
    \scriptsize \emph{Note:} Posterior peak, defined as the response with the largest absolute value over $h=0,\ldots,24$, retaining its sign, at each signed shock size. Responses are cumulated only for variables that enter the model in differences. Boxes give 16--84\% and whiskers 5--95\% posterior credible intervals, and circles are posterior medians. The solid curve is a smooth fit through the medians. The dashed grey line is a linear reference through the medians at the extreme shock sizes. The dotted purple line and the shaded area are the posterior median and the 16--84\% credible interval of the linear version of the model, which has the same identification and the same shock units. Units are percent for industrial production, CPI and the S\&P 500, basis points for the shadow rate and the 10-year yield, percentage points for the excess bond premium, and ratio units for the investment ratio.
    \end{minipage}
\end{figure}

Figure \ref{fig:box_US} depicts the domestic peak responses to a financial shock. For each US variable it shows the posterior distribution of the peak response at each signed shock size. We define the peak as the response with the largest absolute value over horizons $h=0,\ldots,24$, retaining its sign.\footnote{Figure \ref{fig:boxh04} in the appendix repeats the exercise at a fixed horizon of four months. The comparison across shock sizes is then not affected by peaks that occur at different horizons. The output response remains nonlinear in shock size, and its asymmetry again grows with the size of the shock.} The solid curve interpolates the posterior medians. The dashed grey line connects the medians at the two extreme shock sizes. It shows what a model with the same extreme responses, but linearity in shock sizes, would imply. The dotted line and the shaded area show the linear version of the model. Under linearity and sign symmetry, the median responses would lie on a straight line through the origin.

Qualitatively, the responses are in line with the literature \citep[see, e.g.,][]{furlanetto2019identification}. An adverse financial shock raises the excess bond premium and lowers equity prices, industrial production and investment. It also exerts downward pressure on prices and triggers a looser monetary policy stance. Long-term yields decline alongside the policy rate. The new findings are the asymmetries with respect to shock size and sign.

We begin with real activity. For large shocks, US industrial production falls by more than three times as much after an adverse shock as it rises after a benign shock of the same size. For small shocks the adverse response is only moderately larger. The asymmetry thus grows steadily with the size of the shock, which a linear model cannot capture. From two standard deviations onward, the 16--84\% credible intervals of the adverse responses exclude zero. The benign responses are smaller, and their credible intervals include zero at all shock sizes. The output response builds up gradually over several months and then slowly fades.

Financial variables display asymmetries of a similar size. The excess bond premium, equity prices and long-term yields respond much more strongly to adverse than to benign shocks, and so do consumer prices and the policy rate. This is what the model implies when most of the asymmetry arises on impact. The financial transmission function scales the impact of the shock on every variable alike, and propagation adds comparatively little. In our estimates, the asymmetry is thus mainly a property of the financial disturbance itself, and real activity inherits it.

The comparison with the linear benchmark shows how the responses change when transmission is restricted to be linear. The two models share the identification and the shock units, so the gap reflects the nonlinear components of our model. For all US variables the linear responses lie between the two branches of the nonlinear model. On impact, this ordering follows from convexity, since the slope of the linear transmission function is close to the slope of the nonlinear one at zero. The size of the gaps, and how they evolve over the horizon, are not imposed. For shocks of one standard deviation the two models are close. For large shocks, the linear model understates the effects of adverse shocks and overstates those of benign shocks. The second error is much larger than the first. After a large benign shock, the linear model implies an increase of US industrial production that is almost three times as large as in the nonlinear model. For industrial production, consumer prices, the investment ratio, equity prices and the excess bond premium, the linear responses to large benign shocks lie outside the 90\% credible intervals of the nonlinear model. For equity prices, the investment ratio and the excess bond premium, the same holds for large adverse shocks. For consumer prices and the two interest rates, the linear responses to large adverse shocks stay within these intervals.

One further result concerns long-term government bond yields. The 10-year yield falls after an adverse financial shock in all five economies. Its loading on the financial factor is negative everywhere, largest in absolute value for the US and smallest for the euro area. The decline is smallest in Japan, where long rates were held close to zero for much of the sample, and least precisely estimated in the euro area. The responses to benign shocks are imprecisely estimated abroad. An adverse financial shock thus raises the credit spread and pushes safe rates down, which is the usual flight-to-safety pattern. It also generates positive co-movement of long yields across countries, in line with the close correlation of long rates at monthly frequency.

Figures \ref{fig:girfUS} to \ref{fig:girfCA} in the appendix report the full dynamic profiles of each variable for three shock sizes. The profiles are expressed per unit of shock and aligned in sign. Any departure from a single common curve is thus direct evidence of nonlinearity. Figures \ref{fig:irf1} and \ref{fig:irf5} report the responses to shocks of one and five standard deviations.

\subsection{International spillovers of US financial shocks}\label{sec: res_spillovers}

\begin{figure}[!htbp]
    \caption{Peak responses to US financial shocks across countries. \label{fig:box_countries}}
    \centering
    \includegraphics[width=\textwidth]{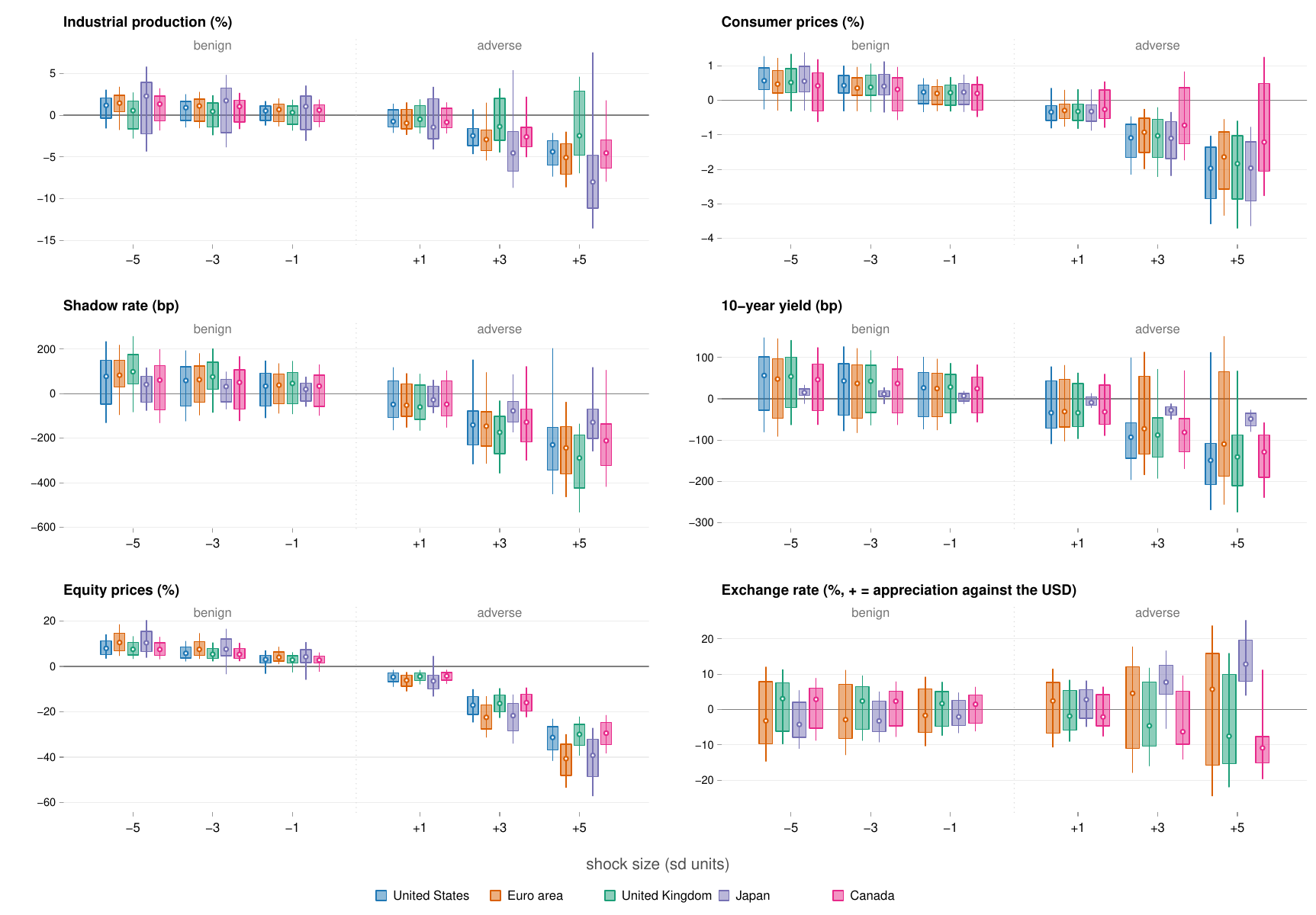}
    \begin{minipage}{\textwidth}
    \vspace{2pt}
    \scriptsize \emph{Note:} Posterior peak, defined as the response with the largest absolute value over $h=0,\ldots,24$, retaining its sign, at each signed shock size. Negative sizes are benign and positive sizes adverse shocks. At each shock size the boxes refer, from left to right, to the US, the euro area, the UK, Japan and Canada. Boxes give 16--84\% and whiskers 5--95\% posterior credible intervals, and circles are posterior medians. Units are percent for industrial production, consumer prices, equity prices and exchange rates, and basis points for shadow rates and 10-year yields. Exchange rates are US dollars per unit of domestic currency, so that a positive response is an appreciation against the dollar.
    \end{minipage}
\end{figure}

This subsection reports the responses of the euro area, the UK, Japan and Canada to the US financial shock. Figure \ref{fig:box_countries} compares the five economies. For each type of variable it shows the posterior distribution of the peak response to shocks of one, three and five standard deviations, with the five countries next to each other. Recall that the impact response of foreign industrial production is restricted to zero, so foreign output can only respond with a lag. Figures \ref{fig:box_EA} to \ref{fig:box_CA} in the appendix report all variables and all shock sizes for each country.

The direction of transmission is the same in all five economies. After an adverse US financial shock, industrial production, consumer prices, policy rates, long-term yields and equity prices fall everywhere. The degree of asymmetry is also similar across countries. In the US, the euro area, Japan and Canada, the peak decline of industrial production after a large adverse shock is more than three times the peak increase after a benign shock of the same size. For shocks of one standard deviation the two responses are much closer. Abroad, the asymmetry thus grows with the size of the shock just as it does in the US.

The size of the output response differs across countries. After a large adverse shock, industrial production in the euro area and in Canada falls by about as much as in the US, and in Japan by considerably more. In all four economies, the 16--84\% credible intervals of the adverse responses exclude zero. The UK is again the exception. The posterior median of UK output declines, but the credible intervals include zero at every shock size. UK industrial production co-moves only weakly with US activity at monthly frequency. Outside the pandemic, its monthly growth rate is essentially uncorrelated with that of US industrial production. The data therefore contain little information about this response.

Consumer prices fall in all five economies after a large adverse shock. Their credible intervals exclude zero everywhere except in Canada. All central banks ease. The shadow rates of the US, the euro area, the UK and Canada fall by similar amounts. The decline in Japan is much smaller, since Japanese policy rates were close to their lower bound for most of the sample. Long-term yields decline everywhere as well, most in the US and least in Japan.

Equity prices react strongly and immediately. A large adverse shock causes sizable losses in all five stock markets, and the adverse response is about four times as large as the benign one in each of them. In standardized units, the loadings of the four foreign indices on the financial factor are close to that of the S\&P 500 (see Figure \ref{fig:lambdaq}). Foreign stock markets thus respond to a US-based financial shock much like the US market. In percent, the EURO STOXX 50 and the Japanese index fall by more than the S\&P 500. This reflects their higher volatility and not a larger loading. The financial factor accounts for most of the contemporaneous variation of all five equity indices. This close co-movement is consistent with the literature on the global financial cycle \citep{rey2013dilemma, miranda2020us}.

The exchange rate responses differ across currencies. After an adverse shock the yen appreciates against the US dollar, in line with its role as a safe-haven currency, while the Canadian dollar depreciates. For large shocks, both responses are clearly different from zero. The responses of the euro and sterling cannot be distinguished from zero.

Figure \ref{fig:profiles} in the appendix reports the response profiles per unit of shock. For most variables and countries, the per-unit response to a shock of five standard deviations is clearly larger in absolute value than that to a shock of one standard deviation. The effects of adverse shocks thus increase more than proportionally with their size, at home and abroad.

\begin{figure}[!htbp]
    \caption{Sign asymmetry of the per-unit response over the horizon. \label{fig:asymhor}}
    \centering
    \includegraphics[width=\textwidth]{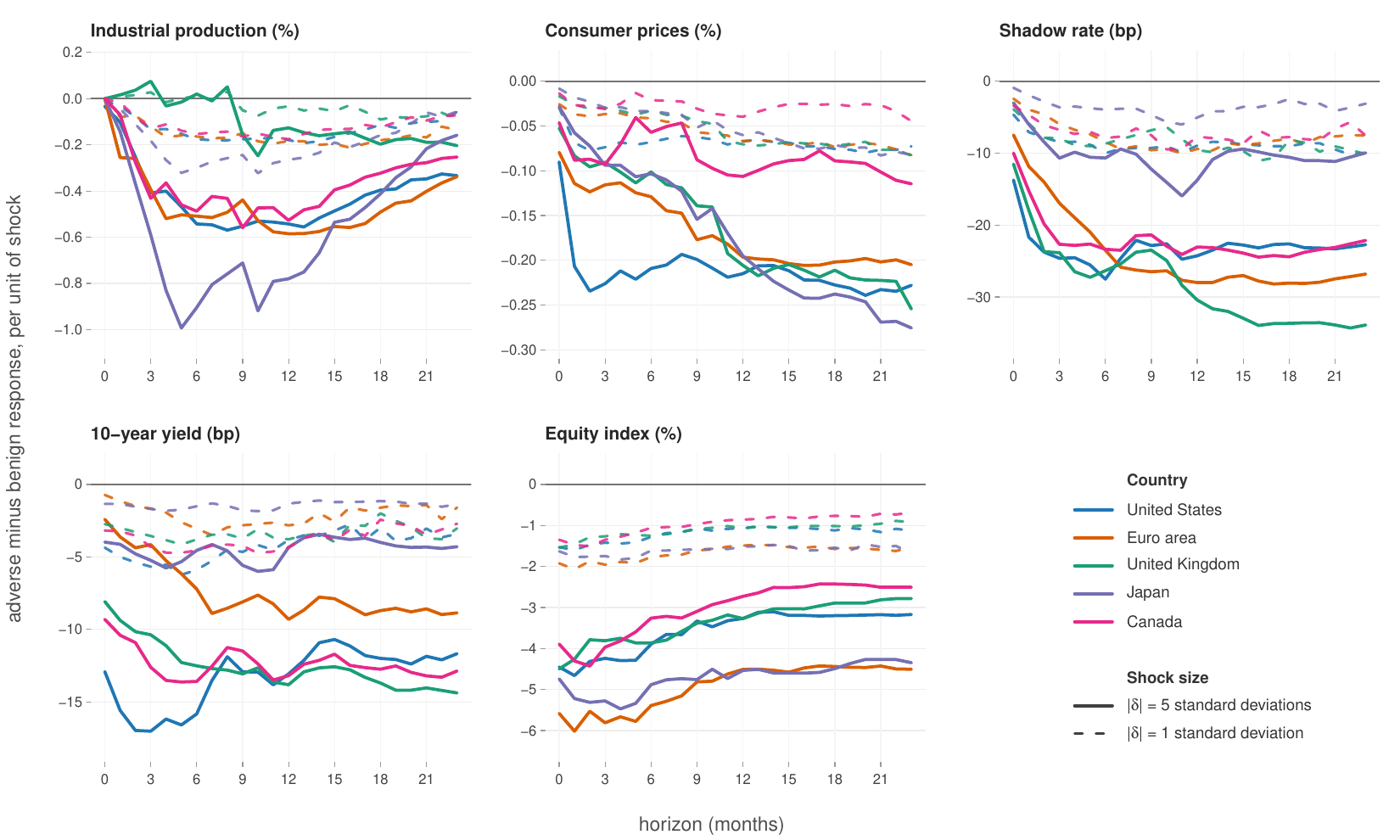}
    \begin{minipage}{\textwidth}
    \vspace{2pt}
    \scriptsize \emph{Note:} For each variable and horizon, the figure plots the per-unit response to an adverse shock minus the sign-flipped per-unit response to a benign shock of the same size, so that a negative value means that the adverse response is larger in absolute value. A symmetric model would lie on the zero line at every horizon. Dashed lines are one standard deviation shocks and solid lines five standard deviation shocks.
    \end{minipage}
\end{figure}

Figure \ref{fig:asymhor} considers the sign asymmetry across horizons. For each variable it plots the difference between the per-unit response to an adverse shock and the sign-flipped per-unit response to a benign shock of the same size. A symmetric model would deliver a flat line at zero. For shocks of one standard deviation the lines lie slightly below zero for most variables. For shocks of five standard deviations, the output lines for the US, the euro area, Japan and Canada drop below zero within the first months and stay there. The adverse response is thus larger in absolute value at every horizon, not only at the peak. The lines for prices, policy rates, long yields and equity prices lie below zero from impact onward, and the gaps are of a similar size relative to the responses. For UK output the gap is small and changes sign over the horizon.

The comparison with the linear benchmark points in the same direction as for the US. For most foreign variables, the linear responses again lie between the two branches of the nonlinear model. After large benign shocks, the linear model clearly overstates the rise of industrial production in the euro area, Japan and Canada, and its responses lie outside the 90\% credible intervals of the nonlinear model. The same holds for consumer prices and equity prices in these economies. After large adverse shocks, the linear model understates the decline of industrial production and of equity prices. For equity prices, the linear responses again lie outside the 90\% credible intervals. For UK industrial production both models imply small and imprecise responses.

Our estimates can be compared with the mostly linear evidence in the literature. \citet{abbate2016changing} find that US financial shocks explain a sizable share of the forecast error variance of UK GDP growth in the late 2000s. \citet{eickmeier2015} find significant negative effects of US credit supply deteriorations on UK activity. Our posterior median for UK industrial production agrees with both in sign, but the response is imprecisely estimated. For the euro area our estimates point to clear and asymmetric negative effects.

The four foreign economies differ mainly in their output responses. Real activity in the euro area, Japan and Canada reacts at least as strongly and as asymmetrically as in the US, whereas UK output does not respond clearly. Financial markets do not explain this difference, since all foreign equity markets respond with similar impact loadings and a similar asymmetry. It thus lies in how financial conditions feed through to real activity.

Overall, the effects of US financial shocks on real activity in the euro area, Japan and Canada are as asymmetric as in the US. In all three economies, output falls more than proportionally with the size of an adverse shock. UK output shows no clear response. Equity prices react strongly in all five economies, with an asymmetry of a similar size. A linear version of the model with the same identification misses these patterns. Its largest error concerns benign shocks, whose effects abroad it overstates by a wide margin.

\subsection{Spillovers and country characteristics}\label{sec: xreg}

The four foreign economies differ mainly in the size of their output responses (see Subsection \ref{sec: res_spillovers}). Several mechanisms could account for these differences. The first is financial structure. Euro area firms depend more on bank credit than UK firms, which rely more on market-based finance \citep[see, e.g.,][]{cecchetti1999legal, gambacorta2014financial}. This could make production in the euro area more sensitive to a tightening of global credit conditions. The second is trade exposure. Canada is closely integrated with the US economy, so a fall in US demand weighs directly on Canadian production. The third is the structure of production. Manufacturing accounts for a larger share of the economy in Japan and the euro area than in the UK and Canada, and it is sensitive to both financial conditions and foreign demand. In Japan, the yen appreciates after an adverse shock, which puts further pressure on exporters. The fourth is the room for monetary easing. Japanese policy rates were close to their lower bound for most of the sample, and the Japanese shadow rate falls by much less than the others. Finally, UK industrial production covers a comparatively small part of the UK economy and includes volatile components such as oil and gas extraction. This may explain why it co-moves so little with US activity at monthly frequency. The model does not identify these channels separately.

To see which of these mechanisms line up with our estimates, we relate the spillovers to the four foreign economies to simple measures of each of them, in the spirit of \citet{georgiadis2016determinants}.\footnote{Table \ref{tab:chars} in the appendix lists these characteristics and their sources.}  We measure the size of a spillover by the log of the absolute peak response to an adverse shock of five standard deviations. Since the volatility of the series differs across countries, we also scale this response by the standard deviation of the monthly changes of the series. We measure the asymmetry by the log of the ratio of the absolute peak responses to an adverse and a benign shock of the same size. We regress each measure on one characteristic at a time, standardized across the four economies. When we pool the five types of variables, we add a fixed effect for each type. We run each regression for every posterior draw, so the posterior intervals reflect the uncertainty about the responses. Table \ref{tab:xreg} reports the results. Shocks of three standard deviations give similar results. With four economies, these regressions are descriptive. The characteristics are correlated with each other, and the regressions cannot tell them apart.

\begin{table}[!htbp]
\caption{Country characteristics and the size and asymmetry of spillovers. \label{tab:xreg}}
\centering \footnotesize \setlength{\tabcolsep}{4pt}
\begin{tabular}{lccccc}
\toprule
 & \multicolumn{3}{c}{Size} & \multicolumn{2}{c}{Asymmetry} \\
\cmidrule(lr){2-4} \cmidrule(lr){5-6}
 & Output & Output & All variables & Output & All variables \\
 & (percent) & (scaled) & (scaled) & & \\
\midrule
Goods exports to the US & $-$0.05 & 0.12 & $-$0.02 & 0.02 & $-$0.02 \\
 & [$-$0.23, 0.13] & [$-$0.07, 0.29] & [$-$0.09, 0.05] & [$-$0.22, 0.26] & [$-$0.13, 0.09] \\[3pt]
Manufacturing share & 0.37 & 0.10 & $-$0.04 & 0.05 & 0.03 \\
 & [0.17, 0.56] & [$-$0.10, 0.28] & [$-$0.12, 0.03] & [$-$0.22, 0.34] & [$-$0.08, 0.15] \\[3pt]
Bank credit relative to market cap. & 0.18 & 0.12 & 0.07 & 0.06 & 0.01 \\
 & [$-$0.01, 0.36] & [$-$0.07, 0.29] & [0.00, 0.13] & [$-$0.18, 0.32] & [$-$0.08, 0.11] \\[3pt]
Months at the lower bound & 0.28 & $-$0.06 & $-$0.08 & $-$0.01 & 0.04 \\
 & [0.12, 0.44] & [$-$0.22, 0.10] & [$-$0.16, 0.00] & [$-$0.24, 0.22] & [$-$0.07, 0.15] \\[3pt]
Co-movement with US output & 0.12 & 0.25 & 0.03 & 0.09 & $-$0.01 \\
 & [$-$0.11, 0.34] & [0.01, 0.46] & [$-$0.04, 0.11] & [$-$0.20, 0.39] & [$-$0.12, 0.10] \\
\midrule
Observations & 4 & 4 & 20 & 4 & 20 \\
\bottomrule
\end{tabular}
\begin{minipage}{\textwidth}
\vspace{4pt}
\scriptsize \emph{Note:} Slopes of separate regressions across the four foreign economies on one standardized characteristic. Posterior medians, with 16--84\% posterior intervals in brackets. Output is industrial production. All variables adds consumer prices, shadow rates, 10-year yields and equity prices.
\end{minipage}
\end{table}

Across these four economies, the estimated output response in percent is positively associated with the manufacturing share. Japan and the euro area have the largest manufacturing sectors and the largest output responses, and the UK has the smallest of both. Bank-based financial systems and long spells at the lower bound also go along with larger output responses, but less closely, and both are correlated with the manufacturing share. Exports to the US are unrelated to the output response. Canada sells far more to the US than the other three economies, but its output does not fall by more than in the euro area. The picture changes once we account for the volatility of industrial production, which is much higher in Japan than elsewhere. Scaled by this volatility, the Japanese output response is no longer the largest, and the manufacturing share no longer explains the differences. The characteristic most closely related to the scaled responses is the co-movement of industrial production with US activity, and this relation is driven largely by the UK. The large Japanese response in percent thus mostly reflects the high volatility of Japanese industrial production. Across all five types of variables, none of the characteristics is clearly related to the size of the scaled responses.

None of the characteristics is related to the asymmetry, neither for industrial production nor for all variables together. All coefficients are small relative to their posterior bands, which include zero. This is in line with how the asymmetry arises in our model. Most of it is present on impact, where it is determined by the transmission function of the financial shock and is common to all variables (see Subsection \ref{sec: res_domestic}). Differences in economic structure thus go along with differences in how strongly a US financial shock spills over to other economies, but not with how asymmetric the spillovers are.

\section{Conclusions}

In this paper, we study asymmetries in the spillovers of US-based financial shocks to the euro area, the UK, Japan and Canada. We introduce a nonlinear factor-augmented multi-country model in which structural shocks enter through univariate transmission functions whose shape can be restricted a priori. We model these functions with monotone Bayesian additive regression trees, and we account for the COVID-19 period through a common volatility scale on the measurement errors. We identify the financial shock through three layers of restrictions. These are sign, zero and point restrictions on the impact loadings, a non-decreasing transmission function for every shock subject to a sign restriction, and convexity of the transmission function of the financial shock. Monotonicity ensures that nonlinear impact responses respect the weak sign restrictions on the loadings. Convexity imposes the direction of financial impact asymmetry documented in the credit shock literature, while its size is estimated.

Our empirical results are threefold. First, the effects of US financial shocks are strongly asymmetric. For large shocks, adverse disturbances move industrial production in the US, the euro area, Japan and Canada by more than three times as much as benign ones of the same size. Second, the asymmetry grows with the size of the shock. For small shocks the two responses are much closer. Third, the asymmetry is broad-based. It is of a similar size for output, prices, interest rates and asset prices, and it is largely present on impact. UK industrial production is the exception, with responses that cannot be distinguished from zero. Across the four foreign economies, differences in the size of the output responses go along with the weight of manufacturing and the volatility of industrial production. The asymmetry is not related to any of the country characteristics we consider.

These results matter for policy assessments of international spillovers. A linear model with the same identification overstates the output expansion following a benign financial shock by a wide margin. It also cannot capture how much more damaging large adverse disturbances are than small ones. Policymakers relying on a linear model that matches the effects of small shocks will thus tend to overestimate the benefits of favorable financial conditions abroad.

Several extensions are possible. The transmission functions are univariate and constant over time. A multivariate version would allow impact asymmetries to differ across variables, and a time-varying shape would allow impact asymmetry itself to depend on the state of the economy. A better measure of monthly UK activity than industrial production would also sharpen the evidence for the UK. The framework could also be applied to a larger set of economies, such as individual euro area member states or emerging markets, where financial heterogeneity is larger than between the aggregates we study here.

\clearpage
\small{\setstretch{0.85}
\addcontentsline{toc}{section}{References}
\bibliographystyle{cit_econometrica.bst}
\bibliography{lit}}

\newpage

\begin{appendices}
\begin{center}
\LARGE\textbf{Appendices}
\end{center}

\setcounter{equation}{0}
\setcounter{table}{0}
\setcounter{figure}{0}
\renewcommand\theequation{A.\arabic{equation}}
\renewcommand\thetable{A.\arabic{table}}
\renewcommand\thefigure{A.\arabic{figure}}

\section{Empirical appendix}\label{App:Emp}

\begin{figure}[!htbp]
    \caption{Responses of industrial production in the four foreign economies to large adverse and large benign US financial shocks. \label{fig:lpcountries}}
    \centering
    \includegraphics[width=\textwidth]{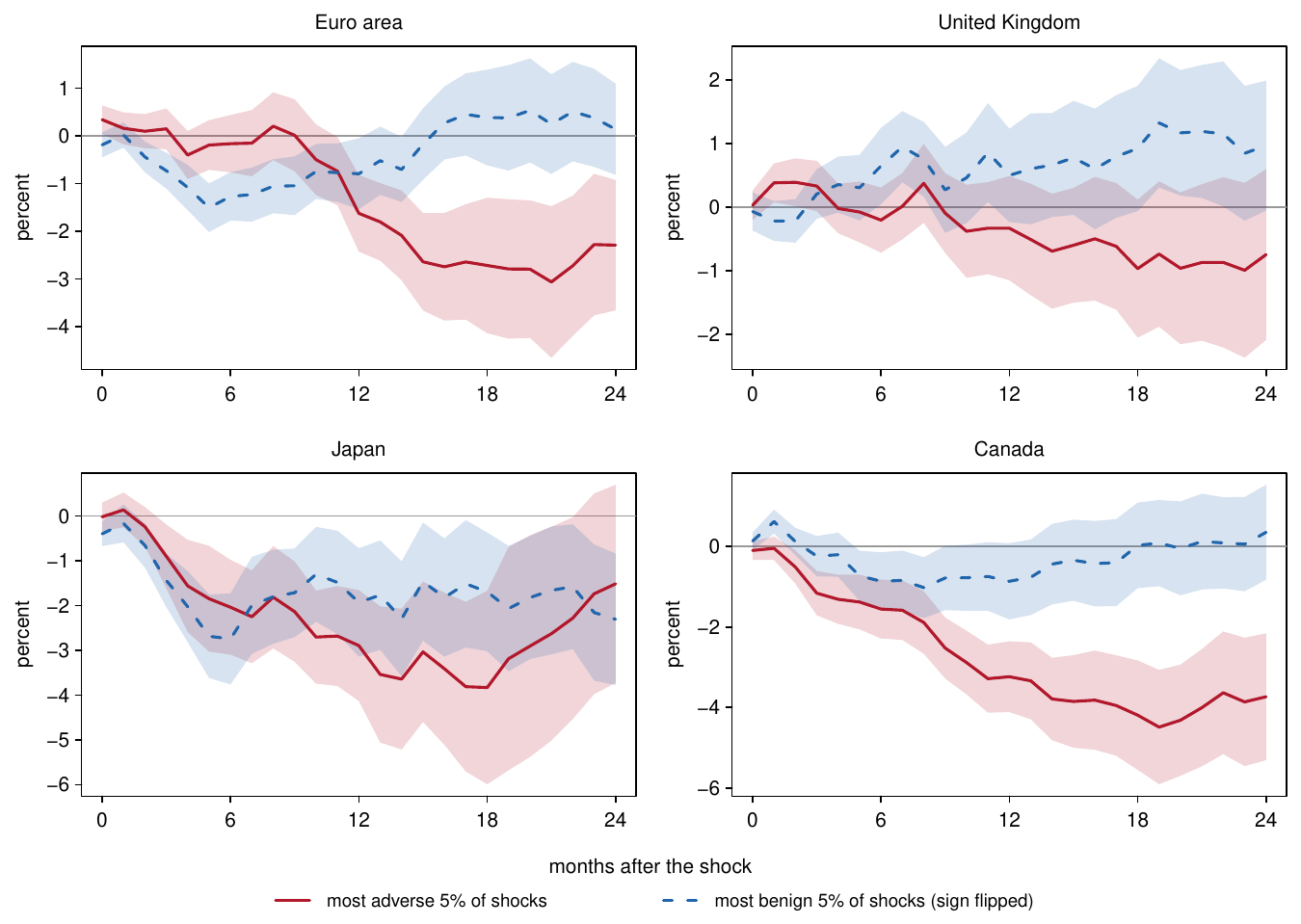}
    \begin{minipage}{\textwidth}
    \vspace{2pt}
    \scriptsize \emph{Note:} See Figure \ref{fig:lp}. Each economy is estimated separately, with the standard errors of \citet{newey1987simple}. The euro area sample starts in 1992M1.
    \end{minipage}
\end{figure}

\begin{figure}[!htbp]
    \caption{Contribution of the identified financial factor to the observables for the EA. \label{fig:hdea}}
    \centering
    \includegraphics[width=\textwidth]{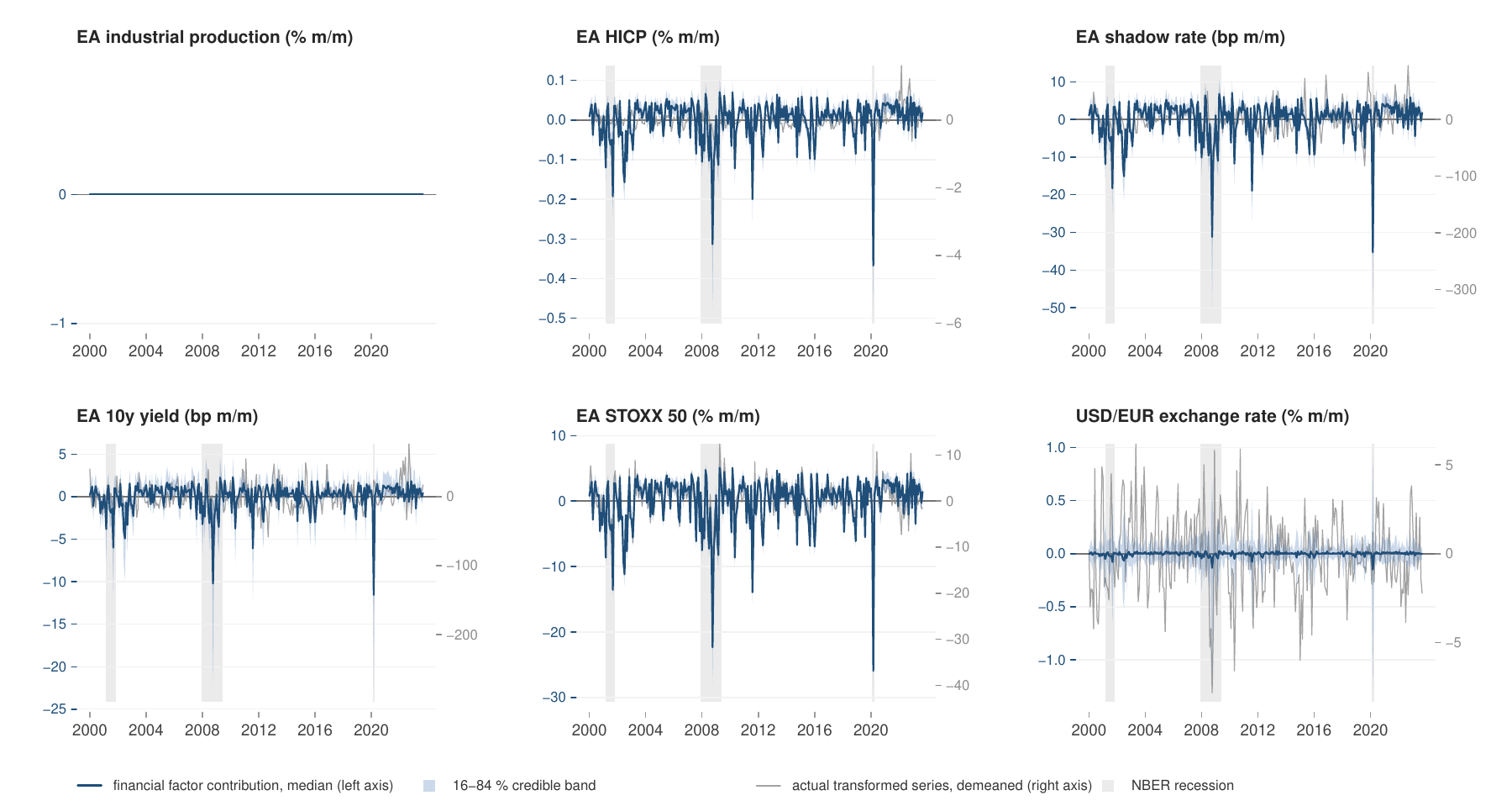}
    \begin{minipage}{\textwidth}
    \vspace{2pt}
    \scriptsize \emph{Note:} See Figure \ref{fig:hd}. Units are monthly percent changes for industrial production, prices, the equity index and the exchange rate, and monthly changes in basis points for the shadow rate and the 10-year yield.
    \end{minipage}
\end{figure}

\begin{figure}[!htbp]
    \caption{Contribution of the identified financial factor to the observables for the UK. \label{fig:hduk}}
    \centering
    \includegraphics[width=\textwidth]{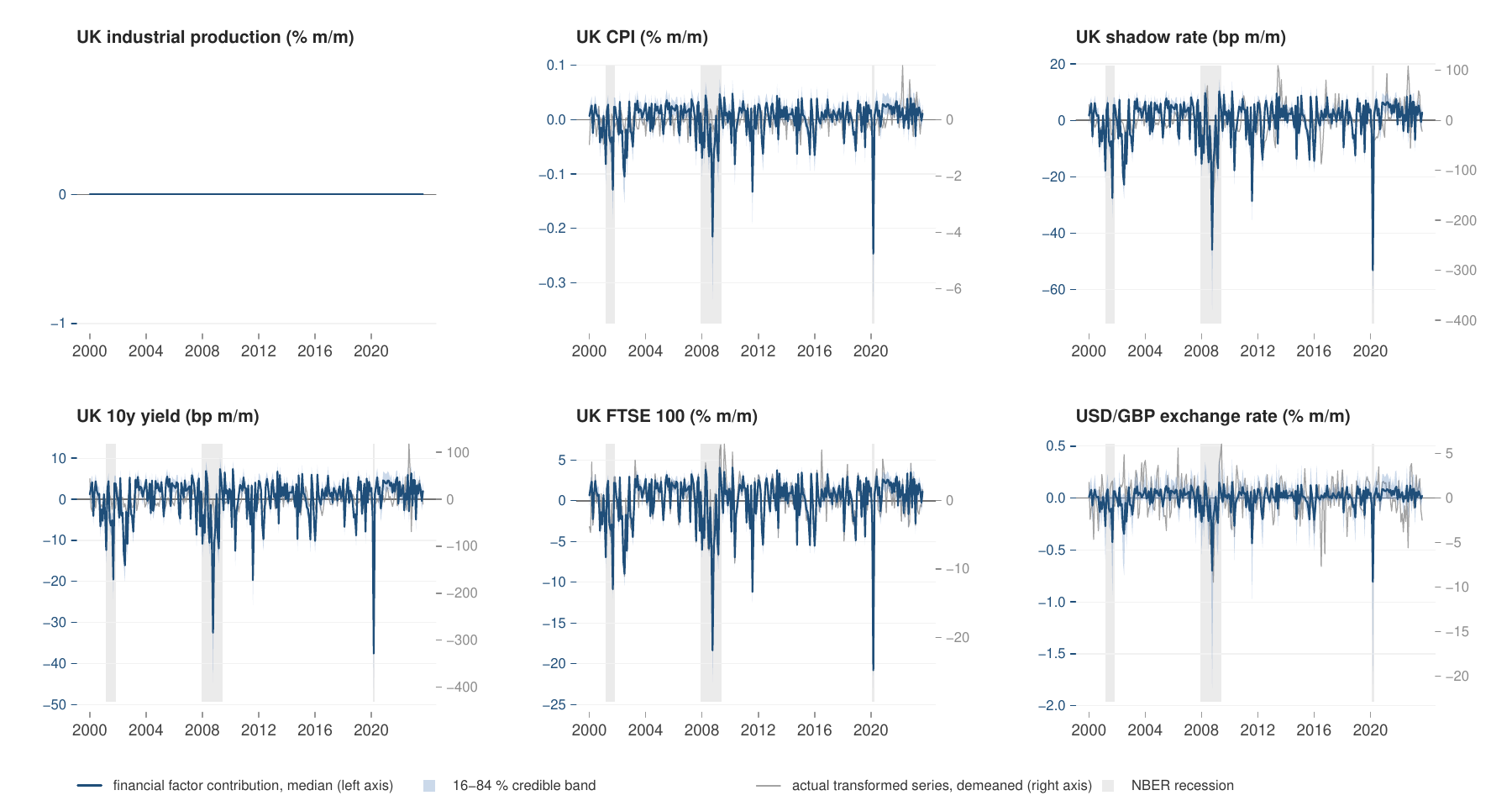}
    \begin{minipage}{\textwidth}
    \vspace{2pt}
    \scriptsize \emph{Note:} See Figure \ref{fig:hd}. Units are monthly percent changes for industrial production, prices, the equity index and the exchange rate, and monthly changes in basis points for the shadow rate and the 10-year yield.
    \end{minipage}
\end{figure}

\begin{figure}[!htbp]
    \caption{Contribution of the identified financial factor to the observables for Japan. \label{fig:hdjp}}
    \centering
    \includegraphics[width=\textwidth]{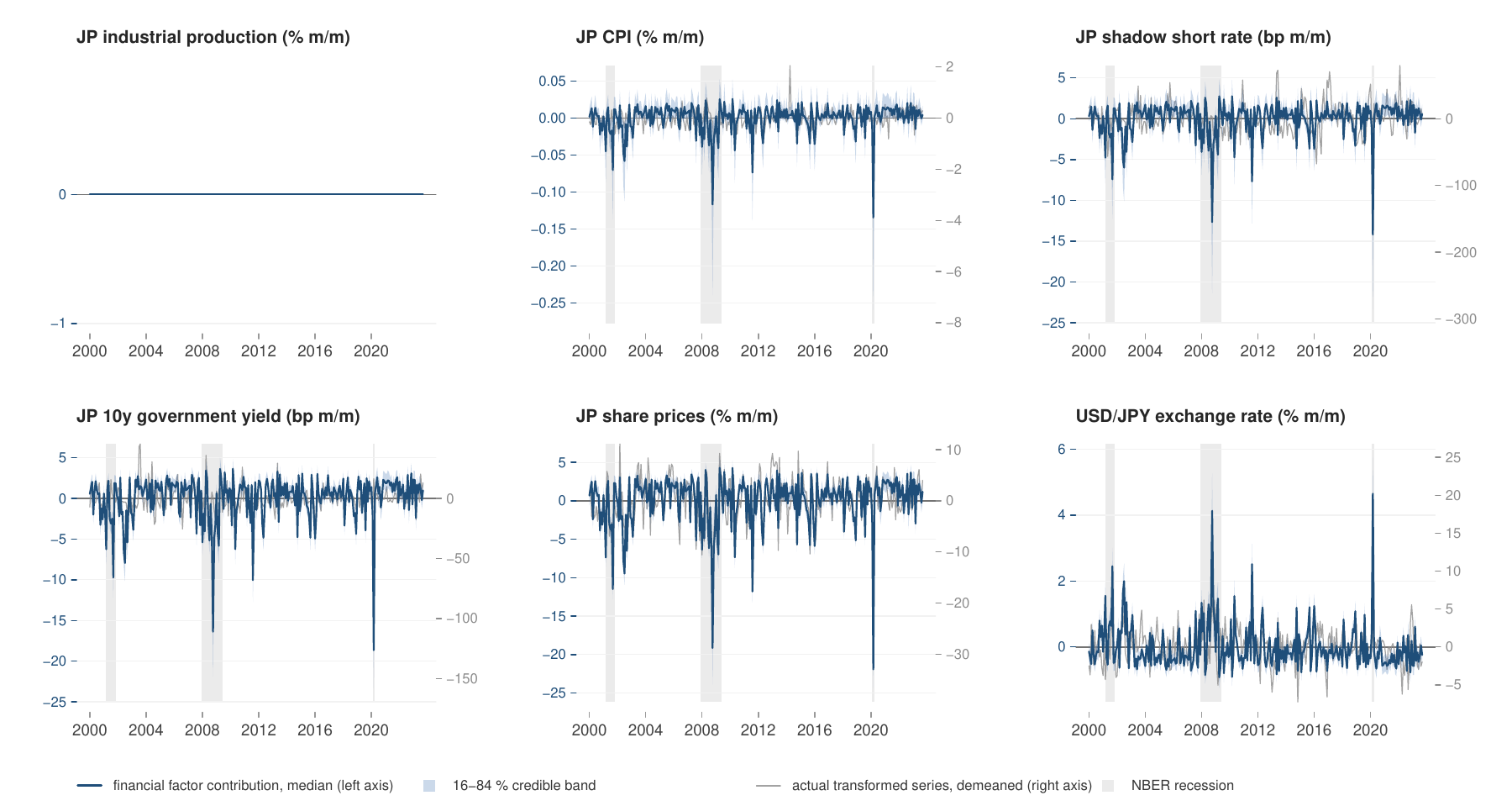}
    \begin{minipage}{\textwidth}
    \vspace{2pt}
    \scriptsize \emph{Note:} See Figure \ref{fig:hdea}.
    \end{minipage}
\end{figure}

\begin{figure}[!htbp]
    \caption{Contribution of the identified financial factor to the observables for Canada. \label{fig:hdca}}
    \centering
    \includegraphics[width=\textwidth]{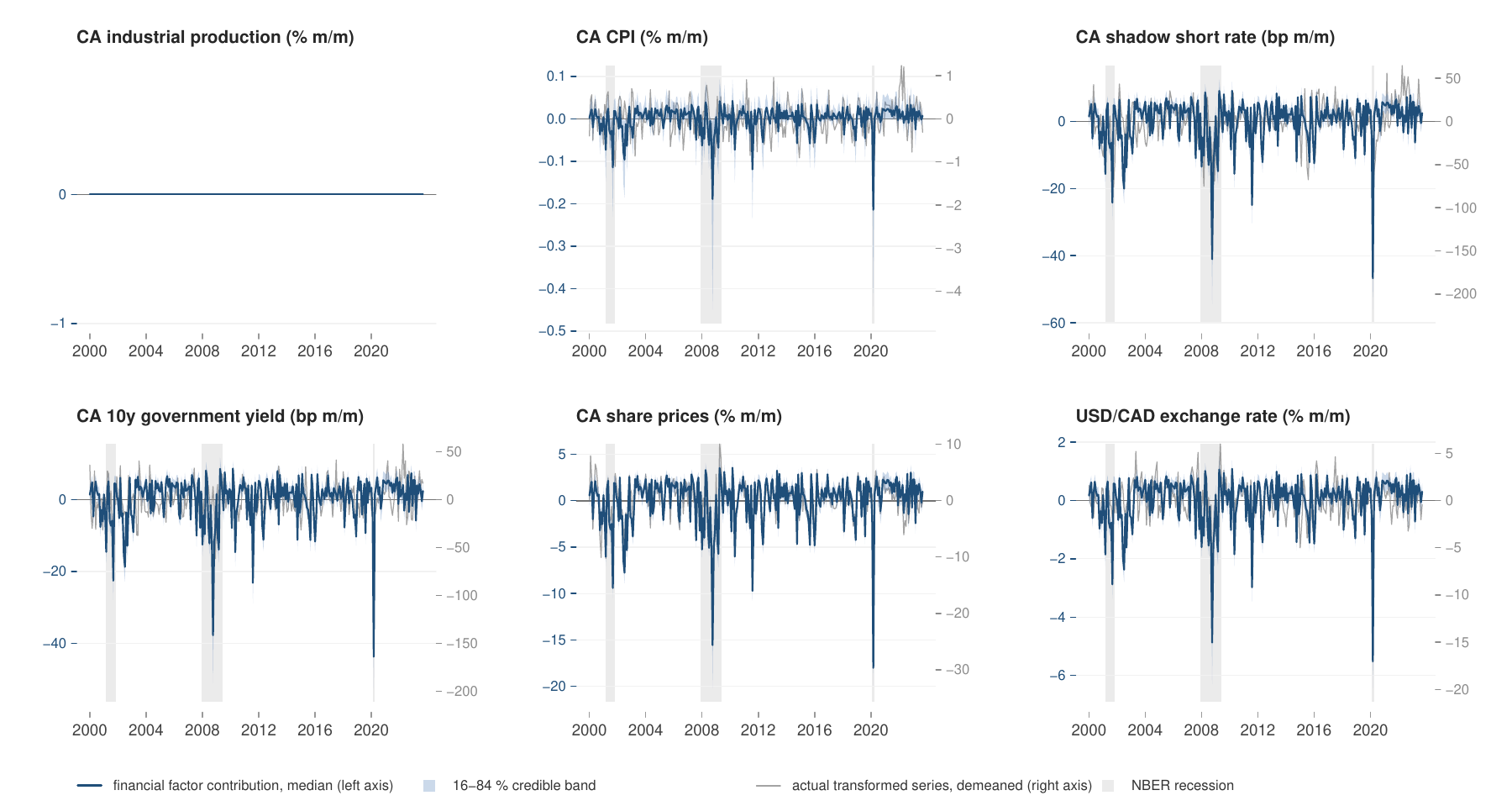}
    \begin{minipage}{\textwidth}
    \vspace{2pt}
    \scriptsize \emph{Note:} See Figure \ref{fig:hdea}.
    \end{minipage}
\end{figure}

\begin{figure}[!htbp]
    \caption{Per-unit, sign-aligned response profiles for the US. \label{fig:girfUS}}
    \centering
    \includegraphics[width=\textwidth]{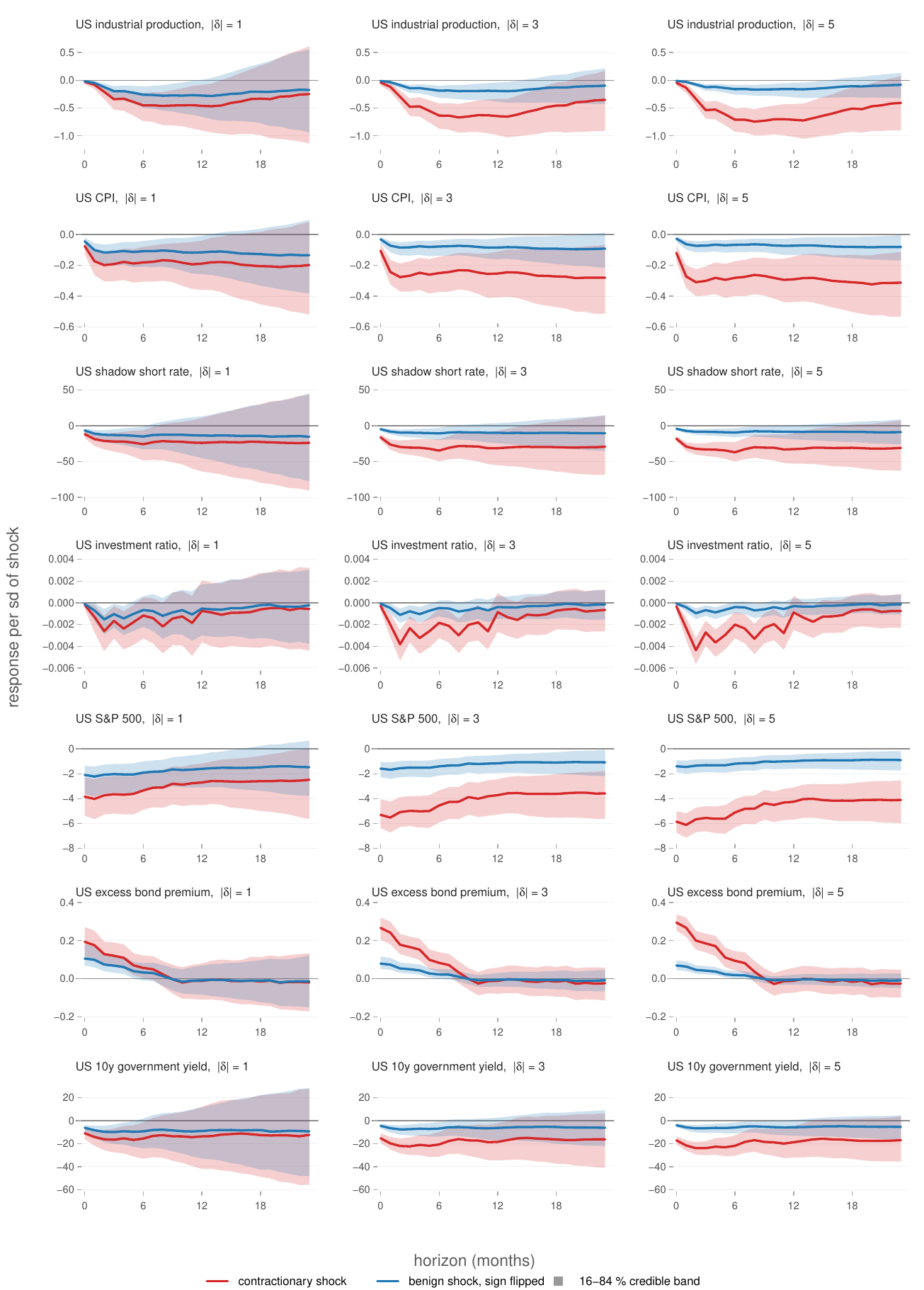}
    \begin{minipage}{\textwidth}
    \vspace{2pt}
    \scriptsize \emph{Note:} Posterior median responses divided by the absolute shock size and aligned in sign, at $|\delta| \in \{1,3,5\}$ standard deviations, with 16--84\% credible bands. The vertical scale is common across the three columns of a row. 
    \end{minipage}
\end{figure}

\begin{figure}[!htbp]
    \caption{Per-unit, sign-aligned response profiles for the EA. \label{fig:girfEA}}
    \centering
    \includegraphics[width=\textwidth]{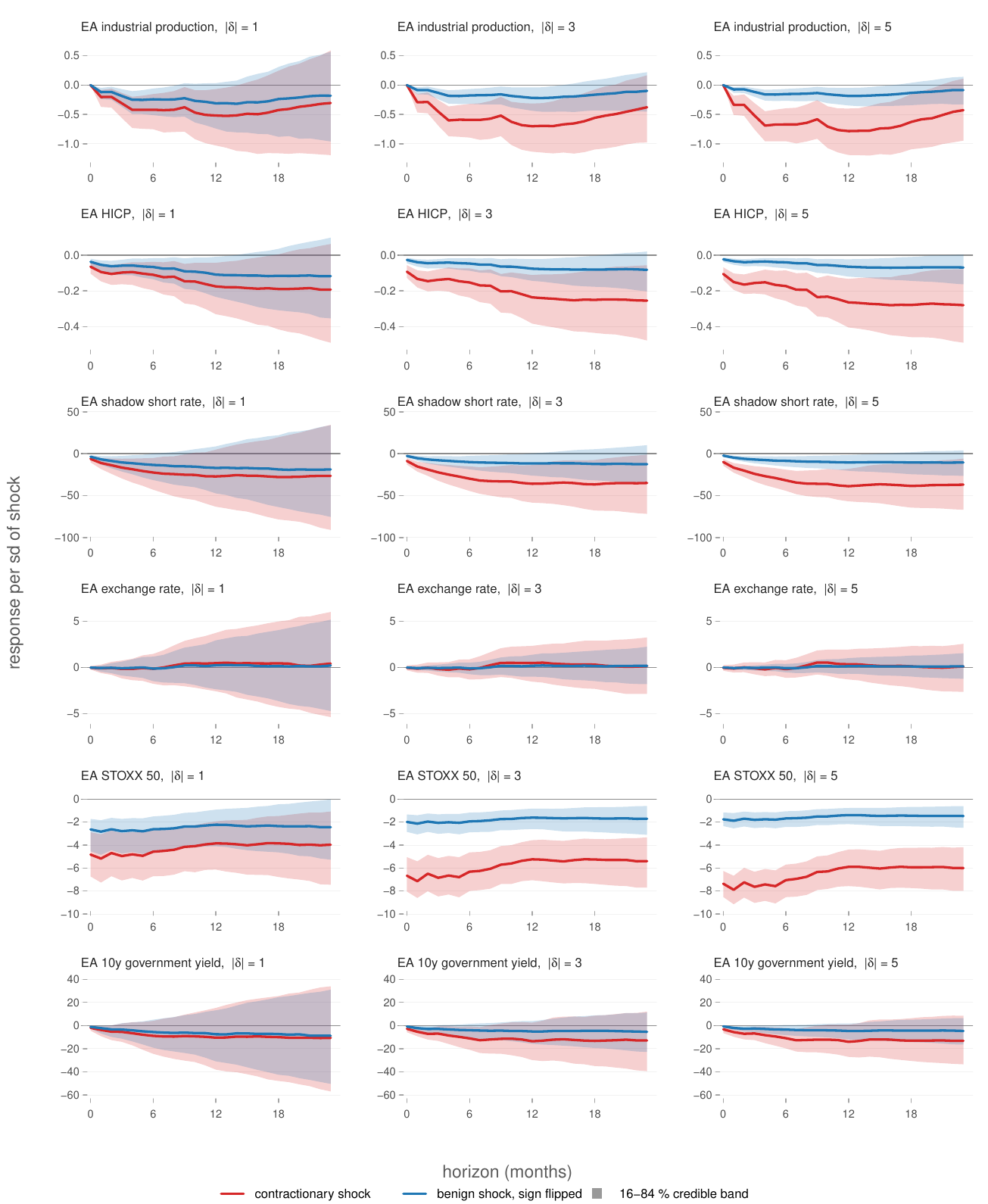}
    \begin{minipage}{\textwidth}
    \vspace{2pt}
    \scriptsize \emph{Note:} See Figure \ref{fig:girfUS}.
    \end{minipage}
\end{figure}

\begin{figure}[!htbp]
    \caption{Per-unit, sign-aligned response profiles for the UK. \label{fig:girfUK}}
    \centering
    \includegraphics[width=\textwidth]{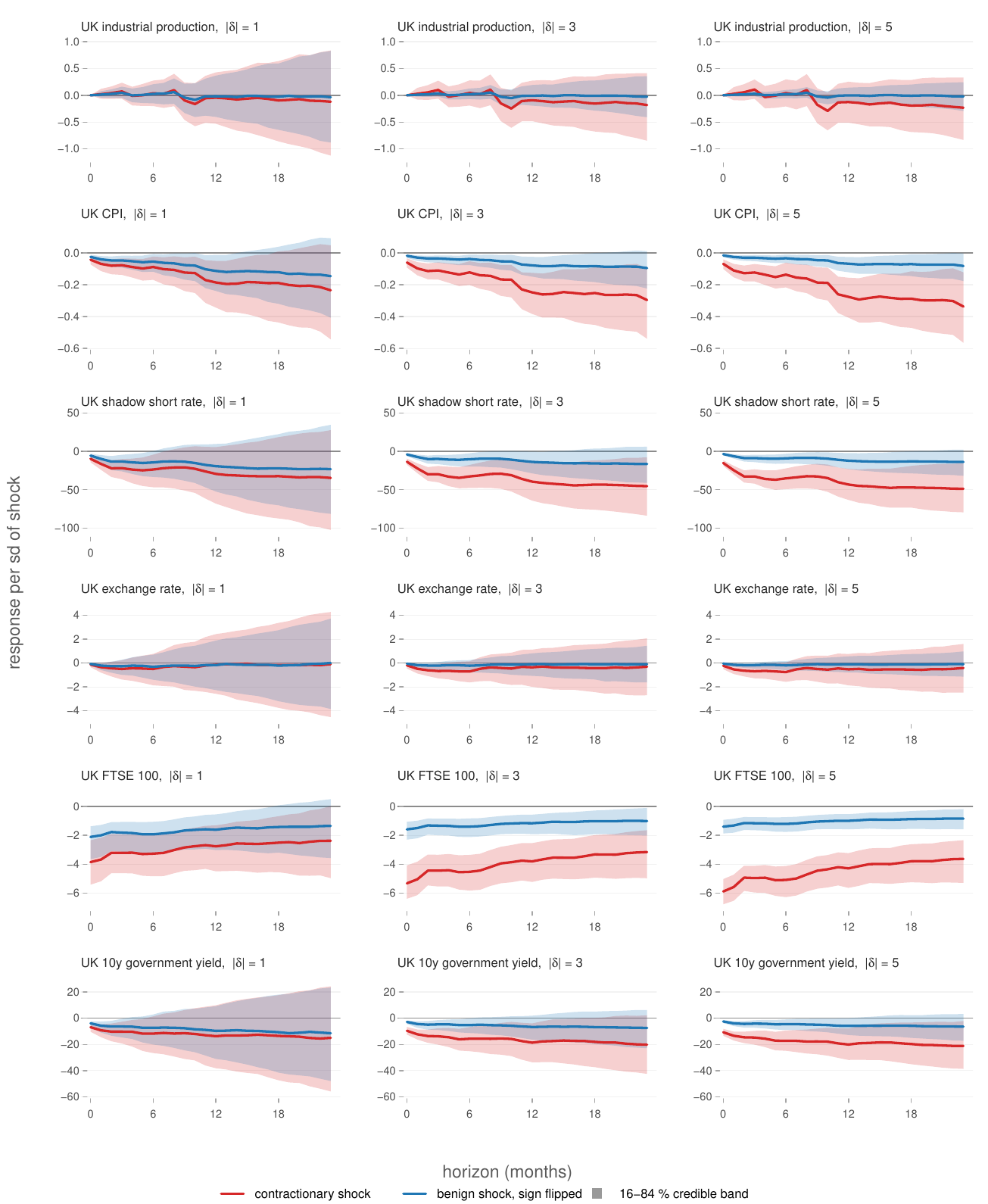}
    \begin{minipage}{\textwidth}
    \vspace{2pt}
    \scriptsize \emph{Note:} See Figure \ref{fig:girfUS}.
    \end{minipage}
\end{figure}

\begin{figure}[!htbp]
    \caption{Per-unit, sign-aligned response profiles for Japan. \label{fig:girfJP}}
    \centering
    \includegraphics[width=\textwidth]{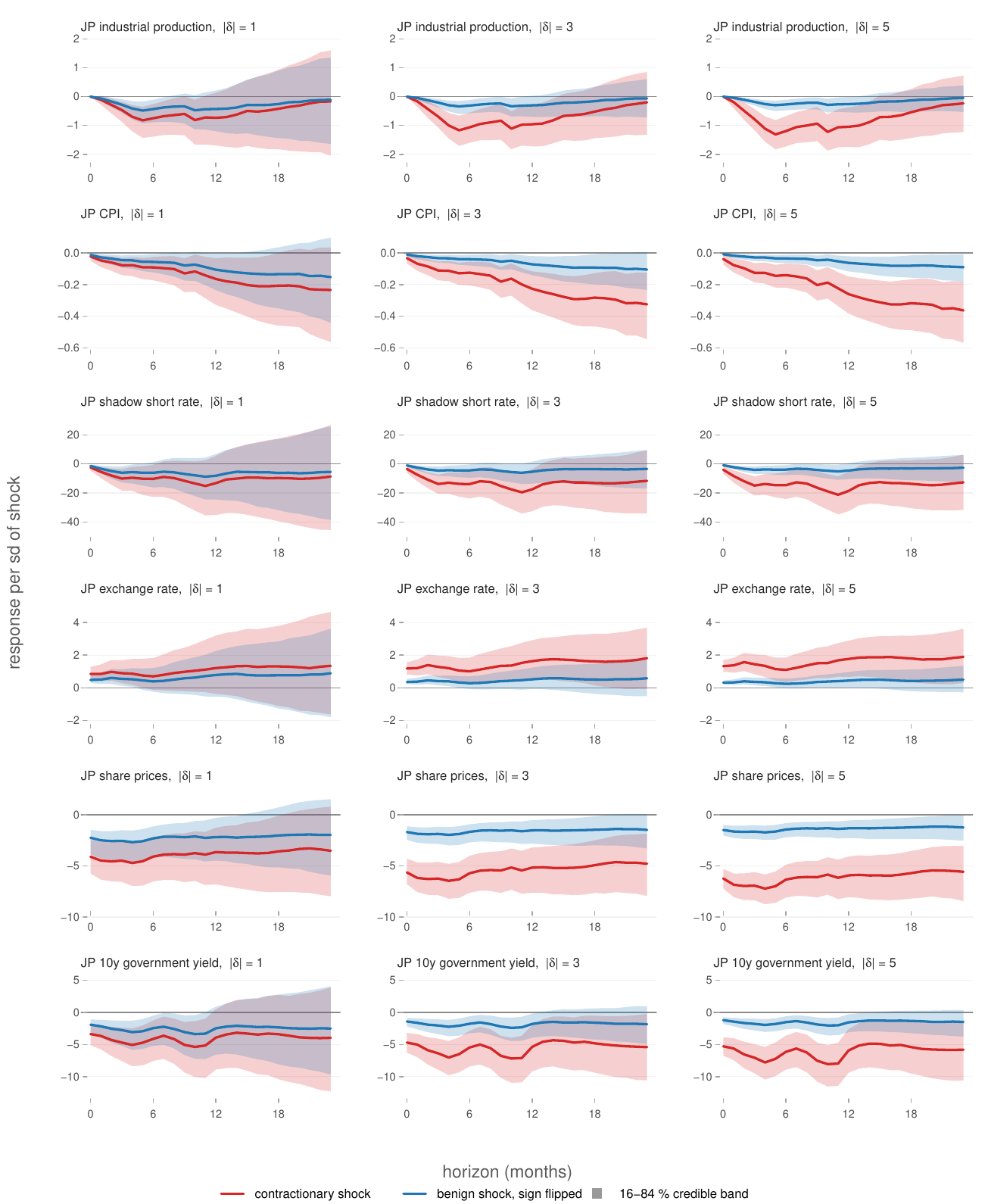}
    \begin{minipage}{\textwidth}
    \vspace{2pt}
    \scriptsize \emph{Note:} See Figure \ref{fig:girfUS}.
    \end{minipage}
\end{figure}

\begin{figure}[!htbp]
    \caption{Per-unit, sign-aligned response profiles for Canada. \label{fig:girfCA}}
    \centering
    \includegraphics[width=\textwidth]{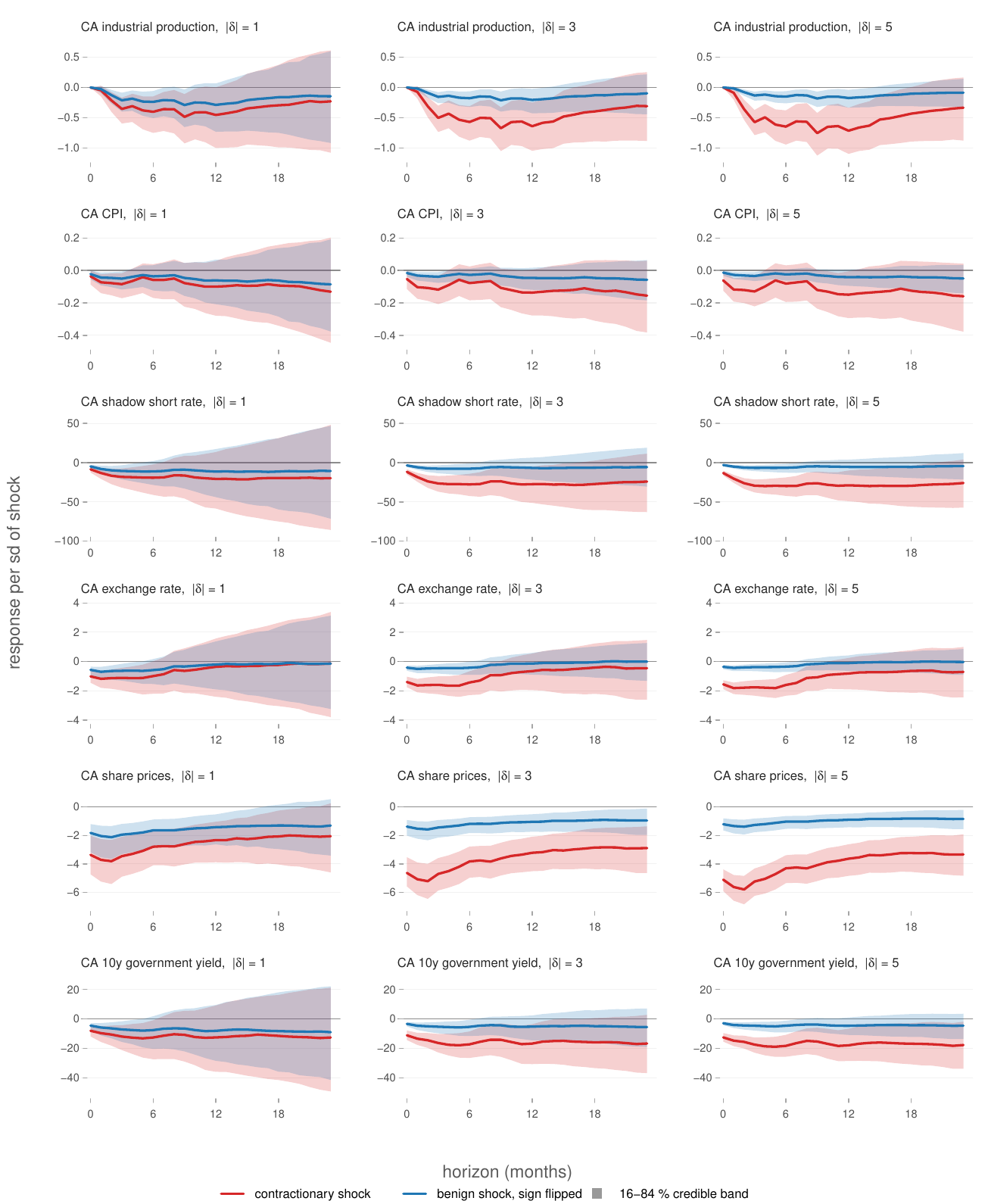}
    \begin{minipage}{\textwidth}
    \vspace{2pt}
    \scriptsize \emph{Note:} See Figure \ref{fig:girfUS}.
    \end{minipage}
\end{figure}

\begin{figure}[!htbp]
    \caption{Peak responses of variables in the euro area to financial shocks in the US. \label{fig:box_EA}}
    \centering
    \includegraphics[width=\textwidth]{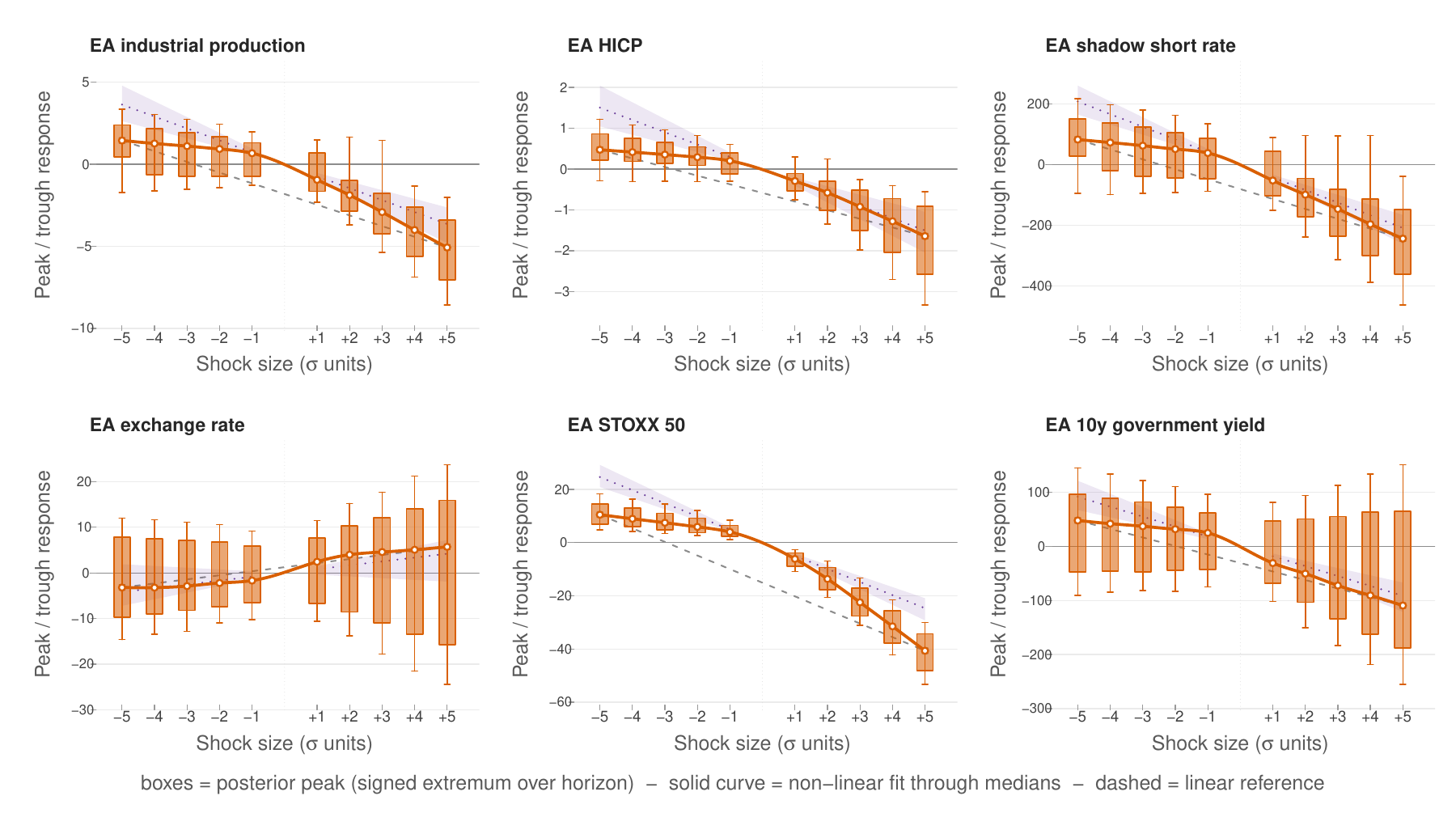}
    \begin{minipage}{\textwidth}
    \vspace{2pt}
    \scriptsize \emph{Note:} See Figure \ref{fig:box_US}. Units are percent for industrial production, HICP, the exchange rate and the EURO STOXX 50, and basis points for the shadow rate and the 10-year yield.
    \end{minipage}
\end{figure}

\begin{figure}[!htbp]
    \caption{Peak responses of variables in the UK to financial shocks in the US. \label{fig:box_UK}}
    \centering
    \includegraphics[width=\textwidth]{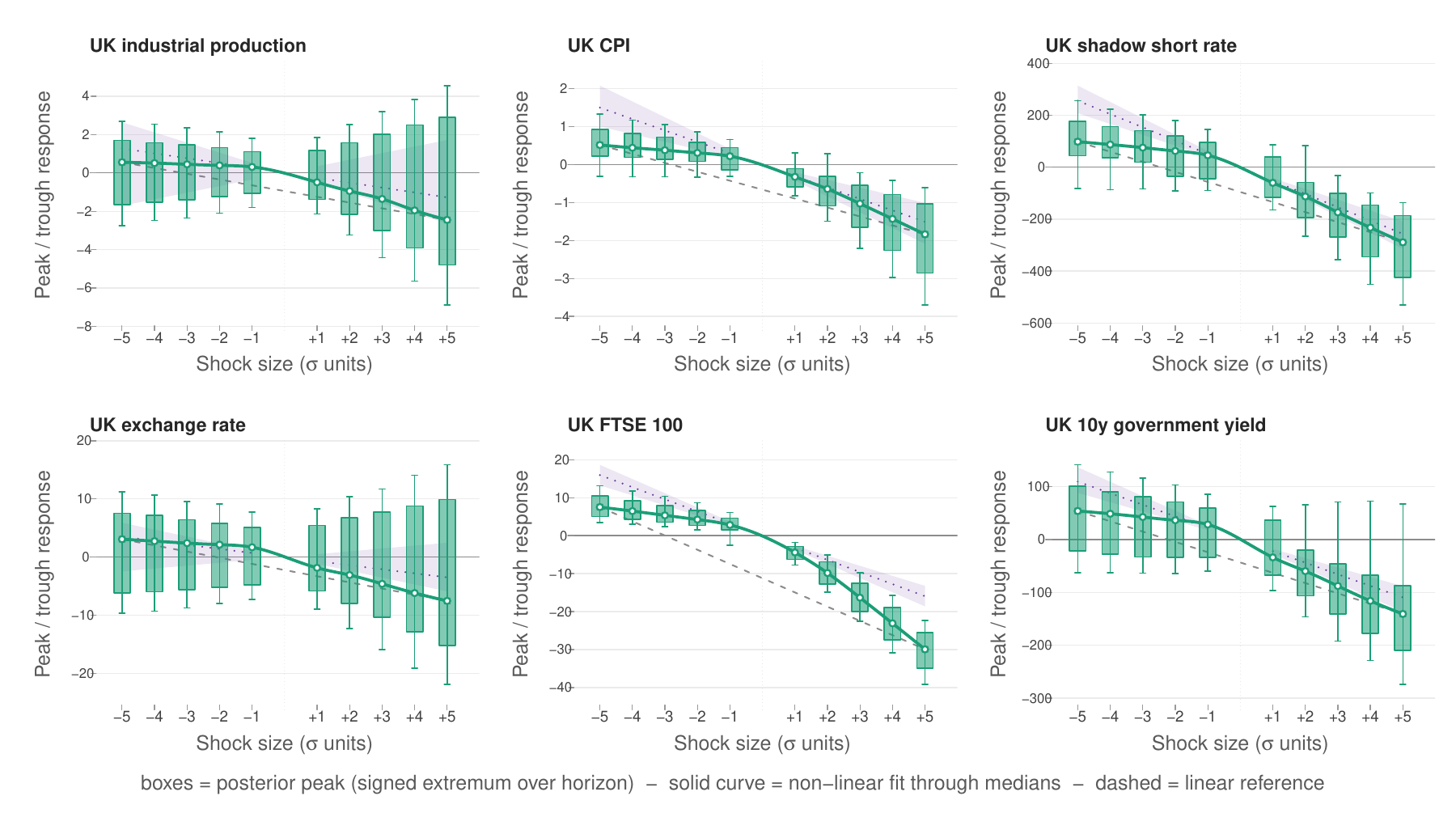}
    \begin{minipage}{\textwidth}
    \vspace{2pt}
    \scriptsize \emph{Note:} See Figure \ref{fig:box_US}. Units are percent for industrial production, CPI, the exchange rate and the FTSE 100, and basis points for the shadow rate and the 10-year yield.
    \end{minipage}
\end{figure}

\begin{figure}[!htbp]
    \caption{Peak responses of variables in Japan to financial shocks in the US. \label{fig:box_JP}}
    \centering
    \includegraphics[width=\textwidth]{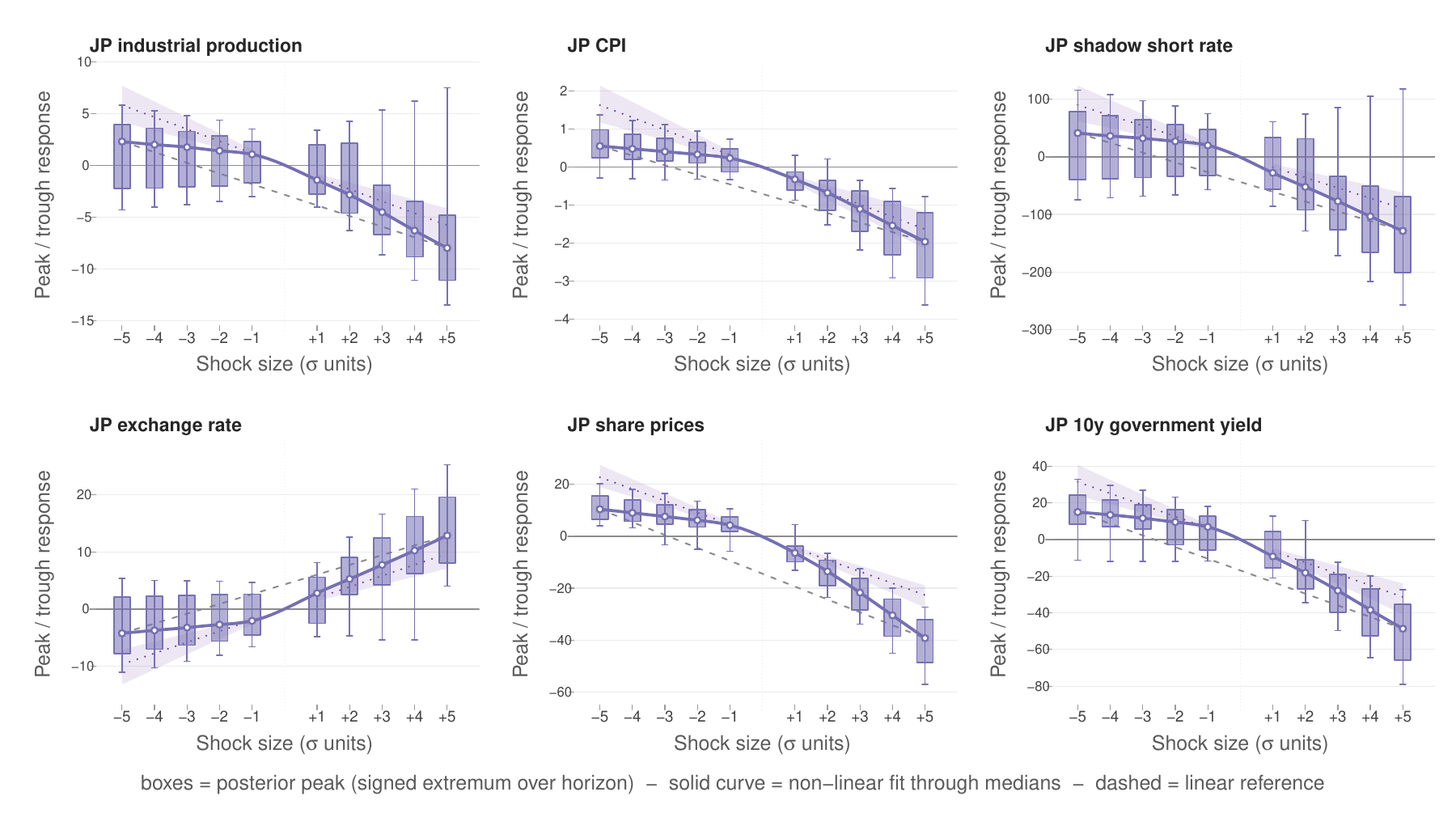}
    \begin{minipage}{\textwidth}
    \vspace{2pt}
    \scriptsize \emph{Note:} See Figure \ref{fig:box_US}. Units are percent for industrial production, CPI, the exchange rate and the share price index, and basis points for the shadow rate and the 10-year yield.
    \end{minipage}
\end{figure}

\begin{figure}[!htbp]
    \caption{Peak responses of variables in Canada to financial shocks in the US. \label{fig:box_CA}}
    \centering
    \includegraphics[width=\textwidth]{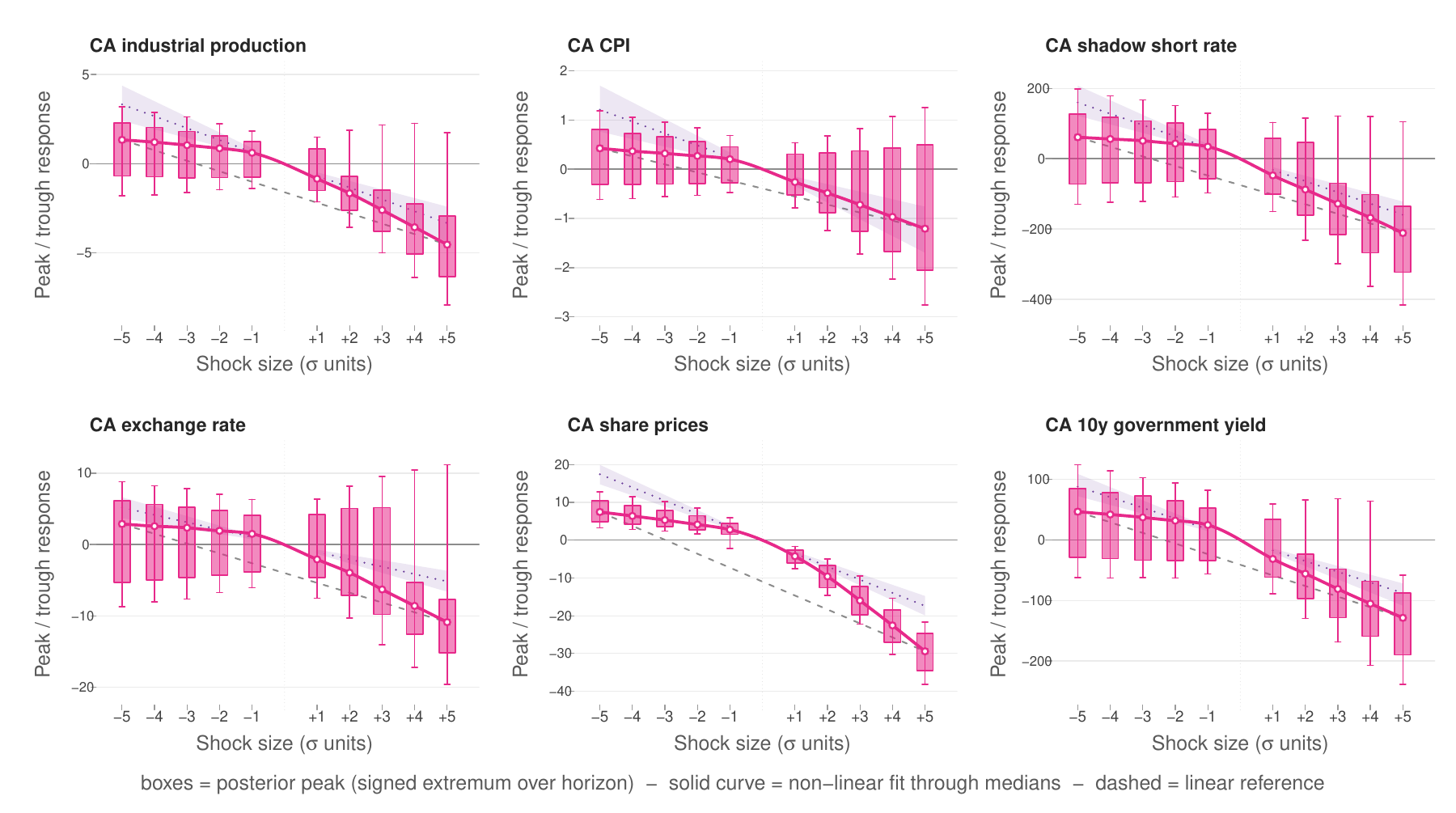}
    \begin{minipage}{\textwidth}
    \vspace{2pt}
    \scriptsize \emph{Note:} See Figure \ref{fig:box_US}. Units are percent for industrial production, CPI, the exchange rate and the share price index, and basis points for the shadow rate and the 10-year yield.
    \end{minipage}
\end{figure}

\begin{figure}[!htbp]
    \caption{International response profiles to a contractionary US financial shock. \label{fig:profiles}}
    \centering
    \includegraphics[width=\textwidth]{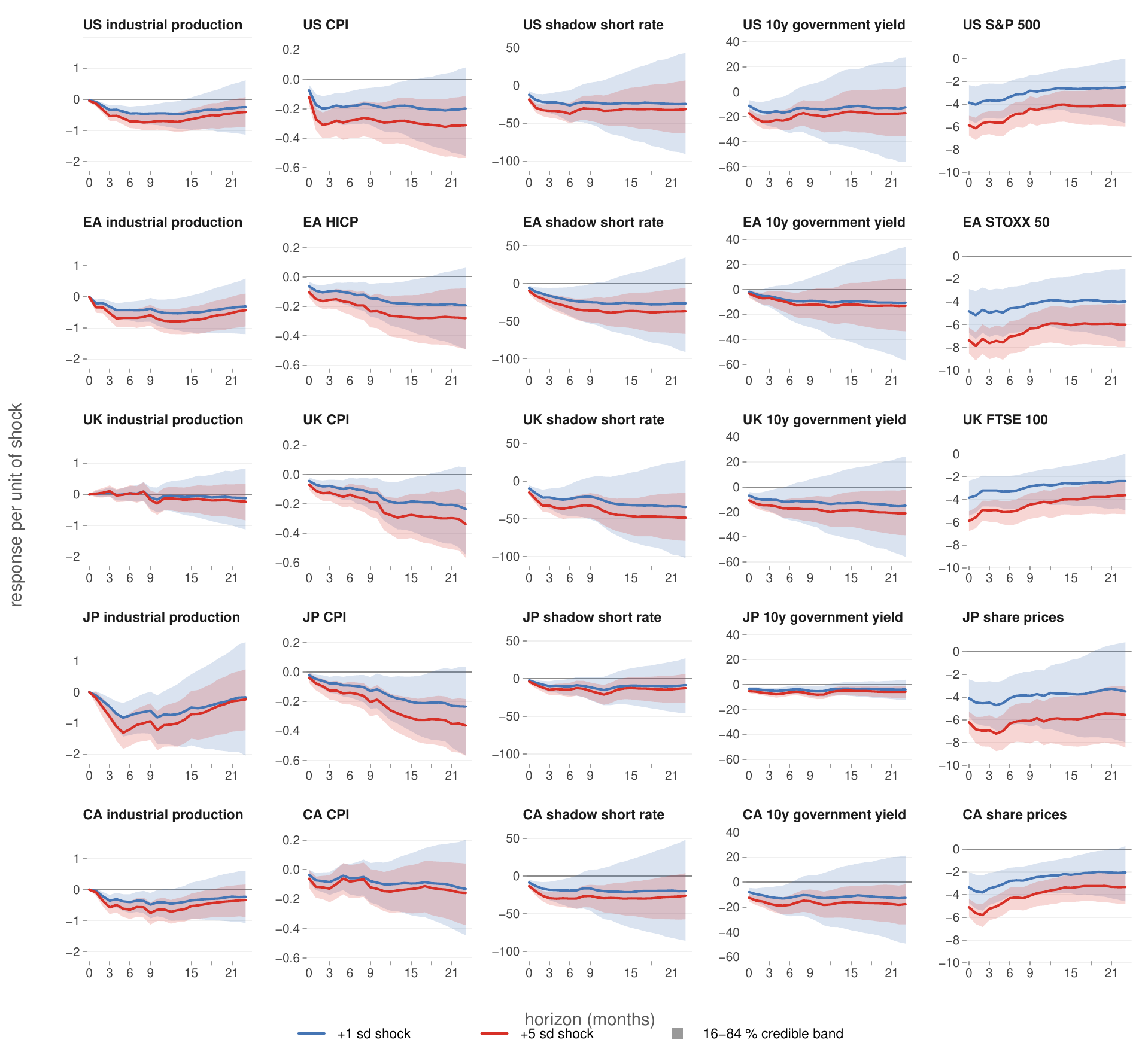}
    \begin{minipage}{\textwidth}
    \vspace{2pt}
    \scriptsize \emph{Note:} Posterior median responses per unit of shock with 16--84\% credible bands. Rows are countries and columns are variable types, with a common vertical scale within each column. Output is industrial production, prices are CPI or HICP, the policy rate is the shadow rate, the long yield is the 10-year government bond yield, and equity is the headline national index.
    \end{minipage}
\end{figure}

\begin{figure}[!htbp]
    \caption{Responses to a one standard deviation contractionary US financial shock. \label{fig:irf1}}
    \centering
    \includegraphics[width=\textwidth]{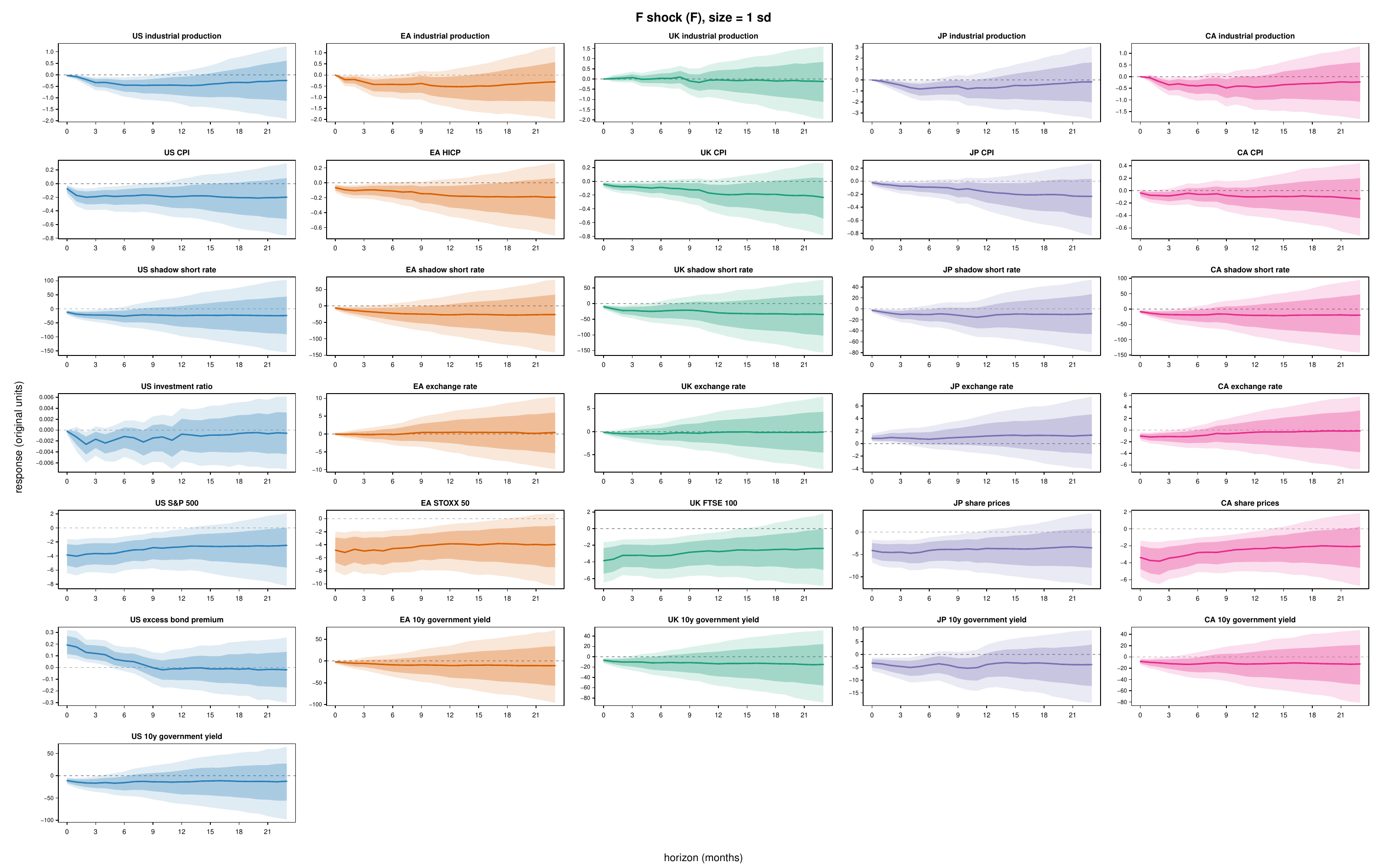}
    \begin{minipage}{\textwidth}
    \vspace{2pt}
    \scriptsize \emph{Note:} Posterior median with 16--84\% and 5--95\% credible bands, for up to $h=24$ months ahead. Units as in Figure \ref{fig:box_US}.
    \end{minipage}
\end{figure}

\begin{figure}[!htbp]
    \caption{Responses to a five standard deviation contractionary US financial shock. \label{fig:irf5}}
    \centering
    \includegraphics[width=\textwidth]{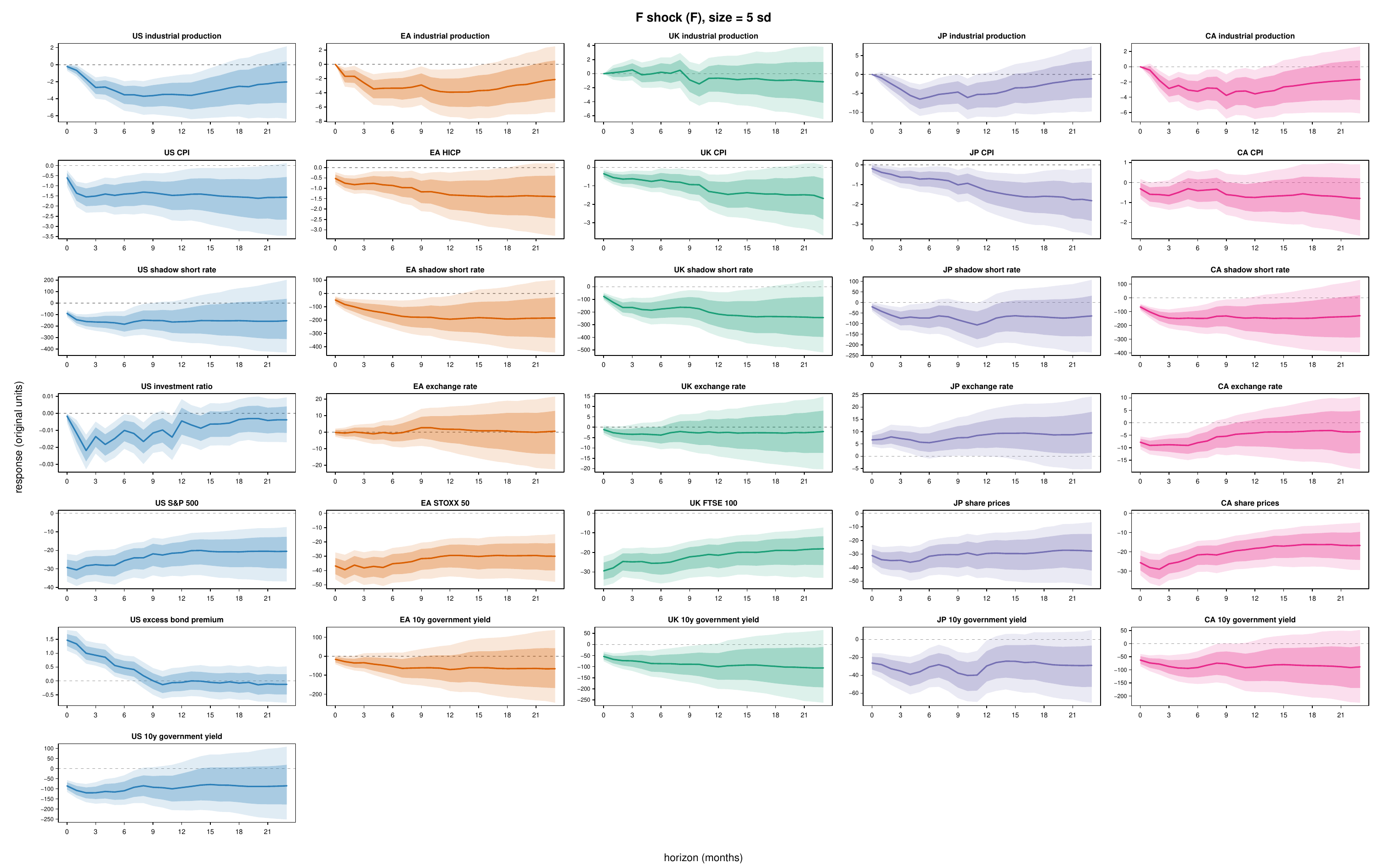}
    \begin{minipage}{\textwidth}
    \vspace{2pt}
    \scriptsize \emph{Note:} See Figure \ref{fig:irf1}.
    \end{minipage}
\end{figure}

\begin{figure}[!htbp]
    \caption{US responses at a fixed horizon of four months. \label{fig:boxh04}}
    \centering
    \includegraphics[width=\textwidth]{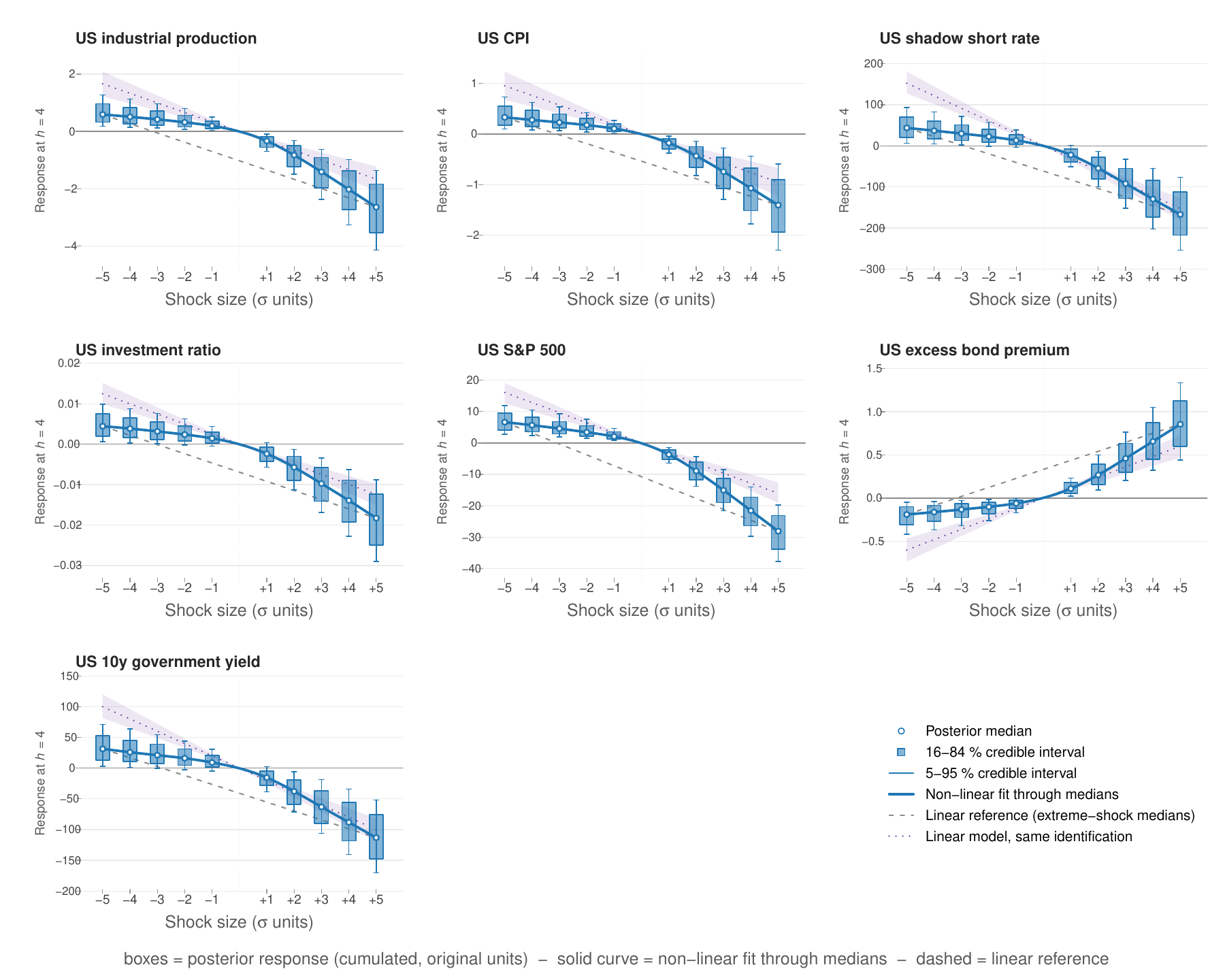}
    \begin{minipage}{\textwidth}
    \vspace{2pt}
    \scriptsize \emph{Note:} As Figure \ref{fig:box_US}, but evaluated at the fixed horizon $h=4$ rather than at the peak, so that the comparison across shock sizes is not affected by the peak occurring at different horizons.
    \end{minipage}
\end{figure}

\begin{table}[!htbp]
\caption{Convergence diagnostics for selected quantities. \label{tab:conv}}
\centering \small
\begin{tabular}{lcc}
\toprule
Quantity & $\hat R$ & ESS \\
\midrule
Slope of $\mu_{q,F}$ at zero & 1.01 & 903 \\
Impact asymmetry of $\mu_{q,F}$, $\delta = 5$ & 1.08 & 42 \\
Loading $\lambda_{mF}$ of the S\&P 500 & 1.10 & 138 \\
Financial shock $q_{F,t}$, 2020M3 & 1.06 & 282 \\
\midrule
Peak response, US industrial production, $\delta = +5$ & 1.01 & 1,274 \\
Peak response, US industrial production, $\delta = -5$ & 1.01 & 1,271 \\
Peak response, EA industrial production, $\delta = +5$ & 1.01 & 1,577 \\
Peak response, UK industrial production, $\delta = +5$ & 1.00 & 1,029 \\
Peak response, JP industrial production, $\delta = +5$ & 1.02 & 1,621 \\
Peak response, CA industrial production, $\delta = +5$ & 1.04 & 1,620 \\
Peak response, excess bond premium, $\delta = +5$ & 1.04 & 593 \\
Peak response, excess bond premium, $\delta = -5$ & 1.03 & 862 \\
Peak response, S\&P 500, $\delta = +5$ & 1.01 & 206 \\
Peak response, S\&P 500, $\delta = -5$ & 1.02 & 138 \\
\midrule
COVID-19 scale $s_t$, 2020M3 & 1.01 & 1,601 \\
COVID-19 scale $s_t$, 2020M4 & 1.01 & 2,703 \\
COVID-19 scale $s_t$, 2020M5 & 1.00 & 764 \\
Decay rate $\rho$ & 1.01 & 529 \\
\bottomrule
\end{tabular}
\begin{minipage}{\textwidth}
\vspace{4pt}
\scriptsize \emph{Note:} $\hat R$ is the potential scale reduction factor of \citet{gelman1992inference}, and ESS is the effective sample size, both computed from the $3{,}000$ retained draws of the three chains. The impact asymmetry of $\mu_{q,F}$ is the ratio of the adverse to the benign increment of the transmission function for a shock of five standard deviations, with both increments averaged over the realized values of the shock.
\end{minipage}
\end{table}

\clearpage
\subsection{Linear benchmark with BART}\label{App:linbart}

\begin{figure}[!htbp]
    \caption{Peak responses of US variables compared with linear transmission functions and BART. \label{fig:boxlinbart}}
    \centering
    \includegraphics[width=\textwidth]{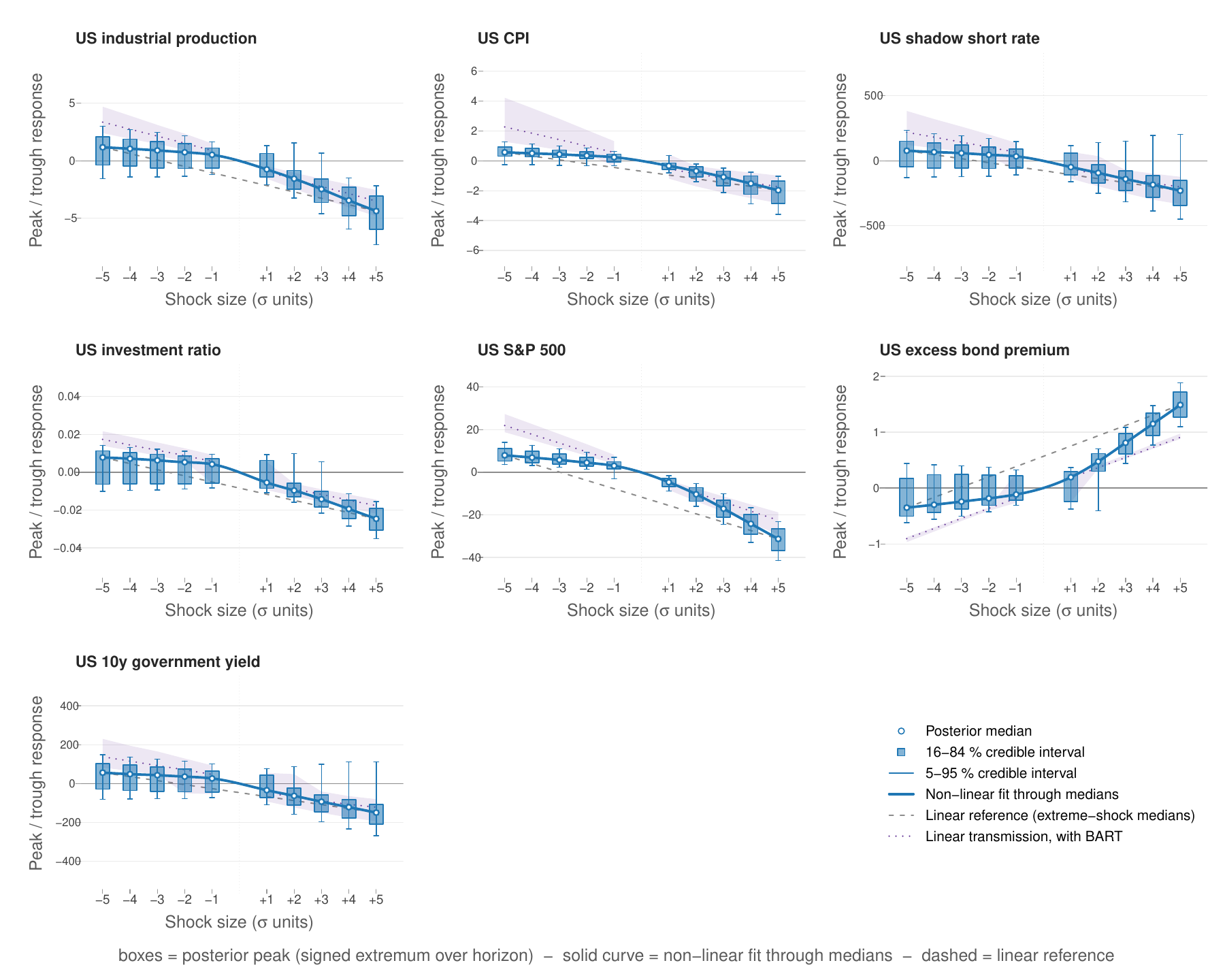}
    \begin{minipage}{\textwidth}
    \vspace{2pt}
    \scriptsize \emph{Note:} As Figure \ref{fig:box_US}, but the dotted purple lines and the shaded areas are the posterior median and the 16--84\% credible interval of a version of our model with linear transmission functions that keeps the BART component. Its peak responses are computed at each shock size.
    \end{minipage}
\end{figure}

Figure \ref{fig:boxlinbart} compares the nonlinear model with a second linear benchmark. It keeps the BART component of our model but sets $\mu_{q,j}(q) = \alpha_j q$. Its responses are close to symmetric in the sign of the shock and close to proportional to its size. They are not exactly so, because the BART component makes propagation state dependent. The benchmark confirms the main finding. It overstates the output effects of large benign shocks by a wide margin. After a large benign shock, US industrial production rises by almost three times as much as in the nonlinear model. After a large adverse shock, it falls by somewhat less than in the nonlinear model, but its response lies within the 90\% credible interval of the nonlinear model. For equity prices and the excess bond premium, the benchmark lies between the two branches of the nonlinear model and outside its 90\% credible intervals at both extreme shock sizes. Unlike the linear benchmark without BART, it implies somewhat larger responses than the nonlinear model also for small shocks, and its credible intervals are wider.

\clearpage
\setcounter{equation}{0}
\setcounter{table}{0}
\setcounter{figure}{0}
\renewcommand\theequation{B.\arabic{equation}}
\renewcommand\thetable{B.\arabic{table}}
\renewcommand\thefigure{B.\arabic{figure}}

\section{Technical appendix}\label{App:Tec}

This appendix summarizes the posterior updates of the sampler described in Subsection \ref{subsec:mcmc}. Throughout, $\bm Y$ is the $T \times M$ matrix of observations, $\bm X$ the $T \times (MP+1)$ matrix of lagged regressors, $\bm\mu$ the $T \times Q_\mu$ matrix of BART factors and $\bm\mu_q$ the $T \times Q_q$ matrix of structural factors, with entries $\mu_{q,j}(q_{j,t})$. The error of variable $m$ in month $t$ has variance $s_t^2 H_m$. Several steps use the same projection. For a loading vector $\bm\varphi$, a column of $\bm\Lambda_\mu$ or $\bm\Lambda_q$, and a partial residual $\bm R_t$ that equals $\bm\varphi f_t$ plus error, let $w = \bm\varphi'\bm H^{-1}\bm\varphi$ and $z_t = \bm\varphi'\bm H^{-1}\bm R_t / w$. Then $z_t = f_t + u_t$ with $u_t \sim \mathcal{N}(0, s_t^2/w)$. The pseudo-observations $z_t$ are the generalized least squares projection of $\bm R_t$ on $\bm\varphi$, and they are a sufficient statistic for $f_t$ given the remaining parameters. One sweep of the sampler cycles through the following steps.

\begin{steps}
\item \emph{VAR coefficients.} Conditional on everything else, the model is a linear regression of $\bm Y - \bm\mu\bm\Lambda_\mu' - \bm\mu_q\bm\Lambda_q'$ on $\bm X$. Equation by equation, after dividing the observations of month $t$ by $s_t\sqrt{H_m}$, the posterior of the $m$th row of $\bm A$ is Gaussian with the usual conjugate moments under the horseshoe prior variance $\bm V_m$.

\item \emph{BART loadings.} The matrix $\bm\Lambda_\mu$ is drawn row by row from a Gaussian posterior obtained by regressing the residual $\bm Y - \bm X\bm A' - \bm\mu_q\bm\Lambda_q'$ on $\bm\mu$, with the same weighting.

\item \emph{BART trees.} For factor $j$, let $\bm\varphi$ be the $j$th column of $\bm\Lambda_\mu$ and let $\bm R_t$ be the residual after removing the VAR part, the structural block and the other BART factors. Each function $\mu_j$ is updated by one Bayesian backfitting step on the pseudo-observations $z_t$, with observation weights $w/s_t^2$.

\item \emph{Structural loadings.} The row $\bm\lambda_m$ of $\bm\Lambda_q$ is drawn from a truncated multivariate Normal on the box $\mathcal{R}_m$. Anchors enter as tight intervals of width $2\varepsilon$ around the target value with $\varepsilon=0.01$, sign restrictions as half lines and zero restrictions as degenerate intervals. Rows without restrictions are drawn from the untruncated Normal.

\item \emph{Structural shocks.} At each date $t$, the shocks are updated jointly with a random walk proposal $\bm q_t^* = \bm q_t + \zeta\bm\xi_t$, $\bm\xi_t \sim \mathcal{N}(\bm 0, \bm I)$, accepted per $t$ with probability $\min\{1, \exp(\Delta\ell_t)\}$. Here $\Delta\ell_t$ is the change in the log posterior, which combines the Gaussian likelihood contribution of $\bm Y_t$, with error variances $s_t^2\bm H$, and the standard Normal prior on $\bm q_t$. The scale $\zeta$ is adapted toward an acceptance rate of $0.234$.

\item \emph{Transmission functions.} For factor $j$, let $\bm\varphi$ be the $j$th column of $\bm\Lambda_q$ and let $\bm R_t$ be the residual after removing the VAR part, the BART factors and the other structural factors. The pseudo-observations $z_t$ then equal $\mu_{q,j}(q_{j,t})$ plus noise with variance $s_t^2/w$. The trees are updated one at a time, each against the residual of all other components. For a tree we propose, with fixed probabilities, to split a terminal node at a grid point (grow), to remove a split whose two children are terminal (prune), or to move such a split to another grid point within its interval (change). The jump at the affected split point is integrated out against its half-Normal prior in closed form, and the move is accepted with the usual MH probability under the tree prior. All jumps of the tree are then drawn one at a time from univariate truncated Normal distributions. The linear coefficient $\alpha_j$ is drawn from a truncated Normal as well. For the financial function, the bound $\dot\mu_{q,F}(0)\geq 0.25$ enters as a lower truncation point for $\alpha_F$ and for the jumps at split points below zero, and moves that would violate it are rejected. For the four non-financial functions the jump scale $\omega_j^2$ is drawn from its inverse-gamma posterior.

\item \emph{COVID-19 scale.} Let $S_t = \sum_m e_{m,t}^2/H_m$ denote the scaled sum of squared errors in month $t$. The likelihood of $s_t$ is proportional to $s_t^{-M}\exp\{-S_t/(2s_t^2)\}$. For March and April 2020, the scale affects only its own month. With the Pareto prior, $s_t^2$ then follows an inverse-gamma distribution truncated to $[1,\infty)$, from which we draw directly. The scale of May 2020 and the decay rate $\rho$ affect all subsequent months, and we update them jointly by a random walk MH step on $(\log \bar s, \log\{\rho/(1-\rho)\})$.

\item \emph{Remaining blocks.} The idiosyncratic variances $H_m$ are updated from inverse-gamma posteriors on the errors divided by $s_t$, and the horseshoe and column-shrinkage hyperparameters from their standard conditionals.
\end{steps}

\clearpage
\setcounter{equation}{0}
\setcounter{table}{0}
\setcounter{figure}{0}
\renewcommand\theequation{C.\arabic{equation}}
\renewcommand\thetable{C.\arabic{table}}
\renewcommand\thefigure{C.\arabic{figure}}

\section{Data appendix}\label{App:Data}

\begin{table}[!htbp]
\caption{Variables, sources and transformations. \label{tab: data}}
\centering \footnotesize \setlength{\tabcolsep}{4pt}
\begin{tabular}{lllc}
\toprule
Variable & Description & Source & Transformation \\
\midrule
US\_INDPRO         & Industrial production           & FRED-MD & $100\Delta\log$ \\
US\_CPI            & Consumer price index            & FRED-MD & $100\Delta\log$ \\
US\_SSR            & Shadow short rate               & \citet{krippner2013ssr} & $100\Delta$ \\
US\_INVESTRATIO    & Investment-to-GDP ratio         & FRED & level \\
US\_SP500          & S\&P 500                        & FRED-MD & $100\Delta\log$ \\
US\_EBP            & Excess bond premium             & \citet{gilchrist2012ebp} & level \\
US\_GovYield10     & 10-year government bond yield   & FRED-MD & $100\Delta$ \\
EA\_INDPRO         & Industrial production           & ECB SDW & $100\Delta\log$ \\
EA\_HICP           & Harmonised index of consumer prices & ECB SDW & $100\Delta\log$ \\
EA\_SSR            & Shadow short rate               & \citet{krippner2013ssr} & $100\Delta$ \\
EA\_EXR            & USD/EUR exchange rate           & FRED & $100\Delta\log$ \\
EA\_STOXX50        & EURO STOXX 50                    & ECB SDW & $100\Delta\log$ \\
EA\_GovYield10     & 10-year government bond yield   & FRED & $100\Delta$ \\
UK\_INDPRO         & Industrial production           & ONS & $100\Delta\log$ \\
UK\_CPI            & Consumer price index            & FRED & $100\Delta\log$ \\
UK\_SSR            & Shadow short rate               & \citet{krippner2013ssr} & $100\Delta$ \\
UK\_EXR            & USD/GBP exchange rate           & FRED-MD & $100\Delta\log$ \\
UK\_FTSE100        & FTSE 100                        & Macrobond & $100\Delta\log$ \\
UK\_GovYield10     & 10-year government bond yield   & FRED & $100\Delta$ \\
JP\_INDPRO & Industrial production & OECD MEI (FRED) & $100\Delta\log$ \\
JP\_CPI & Consumer price index & BIS & $100\Delta\log$ \\
JP\_SSR & Shadow short rate & \citet{krippner2013ssr} & $100\Delta$ \\
JP\_EXR & USD/JPY exchange rate & FRED-MD (inverted) & $100\Delta\log$ \\
JP\_EQUITY & Tokyo Stock Exchange share price index & OECD MEI (FRED) & $100\Delta\log$ \\
JP\_GovYield10 & 10-year government bond yield & OECD MEI (FRED) & $100\Delta$ \\
CA\_INDPRO & Industrial production & OECD MEI (FRED) & $100\Delta\log$ \\
CA\_CPI & Consumer price index & BIS & $100\Delta\log$ \\
CA\_SSR & Shadow short rate & \citet{krippner2013ssr} & $100\Delta$ \\
CA\_EXR & USD/CAD exchange rate & FRED-MD (inverted) & $100\Delta\log$ \\
CA\_EQUITY & S\&P/TSX Composite & OECD MEI (FRED) & $100\Delta\log$ \\
CA\_GovYield10 & 10-year government bond yield & OECD MEI (FRED) & $100\Delta$ \\
\midrule
\multicolumn{4}{l}{\emph{Used for validation only, not included in the estimation:}} \\
US/EA/UK\_CISS     & Composite indicator of systemic stress & \citet{hollo2012ciss} & $\Delta$ \\
\bottomrule
\end{tabular}
\begin{minipage}{\textwidth}
\vspace{4pt}
\scriptsize \emph{Note:} Sample 1999M1--2023M9, monthly. US investment and GDP, which enter the investment ratio, are available at quarterly frequency and are interpolated to monthly frequency following \citet{stock2002macroeconomic}. Impulse responses of variables entering in differences are cumulated to recover level responses, while those of the two variables entering in levels are not. The validation indicators are compared with the financial factor in monthly changes. Exchange rates are US dollars per unit of domestic currency. For Japan and Canada, FRED-MD quotes the exchange rates in domestic currency per US dollar, and we invert them.
\end{minipage}
\end{table}

\begin{table}[!htbp]
\caption{Country characteristics used in Table \ref{tab:xreg}. \label{tab:chars}}
\centering \footnotesize
\begin{tabular}{lcccc}
\toprule
 & Euro area & UK & Japan & Canada \\
\midrule
Goods exports to the US (\% of GDP) & 2.4 & 2.2 & 2.6 & 21.9 \\
Manufacturing value added (\% of GDP) & 15.1 & 9.9 & 20.3 & 11.7 \\
Bank credit relative to stock market capitalization & 1.57 & 0.77 & 1.17 & 0.67 \\
Months at the lower bound (\%) & 39 & 48 & 100 & 23 \\
Co-movement with US industrial production & 0.32 & 0.05 & 0.21 & 0.40 \\
\midrule
Standard deviation of industrial production growth (\%) & 1.07 & 1.20 & 2.10 & 1.00 \\
\bottomrule
\end{tabular}
\begin{minipage}{\textwidth}
\vspace{4pt}
\scriptsize \emph{Note:} Averages over 2000--2023, or over the years available. Goods exports to the US come from the IMF International Trade in Goods statistics (formerly Direction of Trade Statistics) and are divided by GDP in US dollars from the World Bank World Development Indicators (WDI). For the euro area they are the exports of the euro area as a whole. Manufacturing value added and stock market capitalization come from the WDI, and bank credit to the private non-financial sector from the BIS total credit statistics, both in percent of GDP. Months at the lower bound is the share of months from 2000M1 to 2023M9 in which the central bank policy rate (BIS) was at or below 0.5 percent. For Japan, this includes the period from 2013M4 to 2016M9, in which the Bank of Japan had no policy rate target. Co-movement is the correlation of the monthly growth rate of industrial production with that of US industrial production. This correlation and the standard deviation are computed from our data set over 2000M1--2019M12.
\end{minipage}
\end{table}

\end{appendices}

\end{document}